\documentclass[letterpaper,12pt]{article}
\usepackage{amssymb,latexsym,amsmath}

\usepackage{fullpage}
\usepackage{nicefrac}
\usepackage{mathrsfs}%different \cal using \mathscr
\usepackage{slashed}% for dslash

\usepackage[pdftex]{graphicx}
\usepackage[table,dvipsnames]{xcolor}
\usepackage{multirow}
\usepackage{rotating}
\usepackage{bbold}

\makeatletter \@addtoreset{equation}{section}
\makeatother

\begin{document}

\begin{titlepage}
	\thispagestyle{empty}
	\begin{flushright}
		\hfill{DFPD-11/TH/17}\\
		\hfill{ROM2F/2011/18}
	\end{flushright}
	
	\vspace{35pt}
	
	\begin{center}
	    { \LARGE{\bf On the vacua of $N=8$ gauged supergravity \\[3mm]
	    in 4 dimensions}}
		
		\vspace{50pt}
		
		{G.~Dall'Agata$^{1,2}$ and G.~Inverso$^{3,4}$}
		
		\vspace{25pt}
		
		{
		$^1${\it  Dipartimento di Fisica ``Galileo Galilei''\\
		Universit\`a di Padova, Via Marzolo 8, 35131 Padova, Italy}
		
		\vspace{15pt}
		
	    $^2${\it   INFN, Sezione di Padova \\
		Via Marzolo 8, 35131 Padova, Italy}
		
		\vspace{15pt}
		
		$^3${\it Dipartimento di Fisica, Universit\`a di Roma ``Tor Vergata''\\
		Via della Ricerca Scientifica, 00133 Roma, Italy}
		
		\vspace{15pt}
		
		$^4${\it  INFN, Sezione di Roma\\
		Via della Ricerca Scientifica, 00133 Roma, Italy}
		}		
		
		\vspace{40pt}
		
		{ABSTRACT}
	\end{center}
	
	We discuss a simple procedure for finding vacua of gauged supergravity models, based on the variation of the embedding tensor rather than on a direct minimization of the scalar potential. We apply this procedure to $N=8$ gauged supergravity in 4 dimensions. We easily recover many of the previously known vacua, also completing their scalar mass spectrum, and we apply our procedure to find a dozen of new analytical vacuum solutions. The analysis shows an interesting structure on the moduli spaces of these vacua and provides new criteria to determine the expected value of the cosmological constant by a simple inspection of the group properties of the embedding tensor.
	
	\vspace{10pt}

\end{titlepage}

\baselineskip 6 mm

%\date{}

%%%%%%%%%%%%%%%%%%%%%%%%%%%%%%%%%%%%%%%%%%%%%%%%%%%%%%%%%%%%%%
%%%%%%%%%%%%%%%%%%%%%%%%%%%%%%%%%%%%%%%%%%%%%%%%%%%%%%%%%%%%%%

\section{Introduction} % (fold)
\label{sec:introduction}

The maximal supersymmetric gravity theory in 4 dimensions has received a lot of attention since its discovery because of its unique matter content and its special properties.
The hope that it could be used as a basis to construct a unified theory of gauge interactions in the context of a sensible quantum gravity theory has been hampered by the lack of chiral fermions.
However, $N=8$ supergravity is still at the center of current investigations because of its possible perturbative finiteness and because its massive deformations (gauged models) can be related via the gauge/gravity correspondence to the theories of stacks of M2-branes.
These, in turn, could be used as a technical tool to obtain new information on strongly coupled gauge theories related to condensed matter applications.

Another important role of gauged supergravity models is played in the framework of flux compactifications of string theory and especially in the search of metastable vacua with positive energy.
In spite of the fact that $N=8$ models cannot be used to obtain realistic phenomenology, understanding the conditions for the occurrence of de Sitter vacua and for their stability may shed new light on the origin of the difficulties we currently face in order to produce them in the context of low energy string models.

Altogether, a comprehensive classification of all possible massive deformations of $N=8$ supergravity and of their vacua is a desirable and now realistic objective of supergravity analyses.
An important role in this project is played by the embedding-tensor formalism \cite{Nicolai:2000sc}.
Using this formalism, all different gaugings can be described in a single covariant construction that is based on the underlying global symmetry group G of the ungauged theory.
In particular all possible gaugings are encoded in an embedding tensor $\Theta_{M}{}^{\alpha}$ that describes the embedding of the gauge group into the global symmetry group and hence can be characterized group-theoretically.
Actually, $\Theta_{M}{}^{\alpha}$ entirely parametrizes the action of gauged supergravity (up to the choice of symplectic frame) and therefore also its scalar potential and vacua.

In this paper we will show how, by combining the embedding tensor formalism, applied to $N=8$ supergravity in \cite{deWit:2002vt,deWit:2007mt}, and the fact that the scalar manifold is a coset space, the conditions to produce vacua of gauged $N=8$ supergravity can be reduced to a combination of linear and quadratic constraints in the embedding tensor, therefore allowing a simple algebraic identification of many new vacua of the theory.
This realizes an idea proposed in \cite{Inverso} and \cite{Dibitetto:2011gm}.
Although we postpone the discussion of the detailed algorithm to the paper's main body, the idea underlying this analysis is rather simple.
In fact, the allowed gaugings of $N=8$ supergravity can be determined by a series of linear and quadratic constraints on $\Theta$.
Then, among other things, $\Theta$ determines the scalar potential, which depends non-linearly on the 70 scalar fields, but only quadratically on $\Theta$ itself.
Now, since the scalar manifold is the coset E$_{7(7)}/$SU(8), any point can be mapped to the base point by an E$_{7(7)}$ transformation, which is the full duality group of the ungauged theory.
This means that by a proper duality transformation, acting on both the scalar fields and the embedding tensor, we can map any critical point of the scalar potential to the base point of the scalar manifold, at the cost of changing the explicit form of the embedding tensor.
This means that in order to find all the vacua of $N=8$ gauged supergravities one can simply solve the critical point conditions at the base point of E$_{7(7)}/$SU(8), scanning over all allowed embedding tensor values.
Since these appear quadratically in the scalar potential, we reduce the problem of finding extrema to an algebraic problem of solving quadratic constraints on $\Theta$.

We will see that such a simple idea allows us to easily reproduce the known exact vacua as well as to produce new ones as easily.
We believe this is a rather important development in this line of research, since in 30 years only a handful of vacua had been produced \cite{Cremmer:1979uq,deWit:1983gs,Warner:1983vz,Warner:1983du,Hull:1984rt,Hull:1984wa,Hull:1984ea,Hull:1988jw,Hull:2002cv,Fischbacher:2010ec,Dibitetto:2011gm} and only very recently, thanks to new numerical techniques, a systematic search has been initiated \cite{Fischbacher:2009cj,Fischbacher:2010ki,Fischbacher:2011jx}.
Actually, a combination of the technique proposed here with the improved numerical analyses of \cite{Fischbacher:2009cj,Fischbacher:2010ki,Fischbacher:2011jx} could allow for a fast reaching of the goal of an exhaustive study of the vacua.

As an intermediate step one could try to obtain a complete classification of the vacua with specific gauge groups.
Unfortunately, as we will see, although the conditions are quadratic, one may still need to resort to numerics for a complete scan.
For this reason, in this paper we will concentrate on few specific classes of gaugings, mainly those realized as subgroups of the electric SL(8) group, 
for which a map between gaugings and geometric and non-geometric fluxes is well established \cite{D'Auria:2005rv,Dall'Agata:2007sr,companion}.
Our discussion is also going to be carried out completely within the framework of 4 dimensional models, but a discussion of the possible embeddings in M-theory for some of the models treated here can be found in a companion paper \cite{companion}.

% section introduction (end)

\section{$N=8$ supergravity} % (fold)
\label{sec:n_8_supergravity}

Before discussing the details of the minimization procedure and the vacua we obtained, we need to fix our notations and conventions.
For this reason in this section we present some technical details about $N=8$ supergravity such as the parameterization of the scalar manifold as well as of the duality group generators and a discussion of the key elements of the embedding tensor approach to gauging supergravity theories \cite{LPTENS 79/6,deWit:2007mt}.

The $N=8$ supergravity multiplet contains 28 vector fields and 70 scalars, parameterizing the coset manifold E$_{7(7)}$/SU(8).
The 28 vector fields can be used to make local a subgroup of the isometries of the scalar manifold\footnote{More precisely, they can gauge a group whose adjoint representation is embedded in the duality group generated by the isometries of the scalar manifold.
This is in general a quotient of the one generated by the representation of the vector fields with its maximal abelian ideal (see section 2.3 of \cite{Dall'Agata:2007sr}).} and, by the gauging procedure, generate a scalar potential, which is the main object of interest in our analysis.

In order to be specific, we need to fix a parameterization for both the generators of the duality group and for the scalar manifold.
A simple way to describe the E$_{7(7)}$ generators follows by decomposing its fundamental and adjoint representations with respect to an ${\mathfrak sl}(8,{\mathbb R}$) subalgebra:
\begin{equation}\label{sl8 decomposition}
 {\bf 56} \longrightarrow {\bf 28} + {\bf 28'},\quad
 {\bf 133} \longrightarrow {\bf 63} + {\bf 70}.
\end{equation}
${\mathfrak e}_{7(7)}$ generators are expressed in terms of $\mathbf{63}$ ${\mathfrak sl}(8,{\mathbb R}$) matrices $\Lambda$  and $\mathbf{70}$ (selfdual and anti-selfdual) real forms $\Sigma_{ABCD}$:
\begin{equation}\label{e7 gen in sl8 basis}
  [t_\alpha]_{M}{}^{N} \equiv \begin{pmatrix}
      \Lambda_{AB}{}^{CD} & \Sigma_{ABCD} \\
      \star\Sigma^{ABCD} & \Lambda'^{AB}{}_{CD} \\
  \end{pmatrix},
\end{equation}%
Here $\Lambda_{AB}{}^{CD} \equiv 2 \Lambda_{[A}{}^{[C} \delta_{B]}{}^{D]}$ and
$\Lambda' = -\Lambda^T$, with  $A,B,\ldots$ denoting the $\bf8$ and $\bf8'$ representations of SL(8,${\mathbb R}$), the actual maximal subgroup of E$_{7(7)}$ being SL(8,$\mathbb R$)$/\mathbb{Z}_2$.
Given the structure of the scalar manifold, it is also useful to rewrite the same generators in a complex basis in which the maximal compact subgroup SU(8)$/\mathbb{Z}_2$ is manifest. 
In this case one obtains a decomposition of the E$_{7(7)}$ generators in terms of the $\mathbf{63}$ generators of SU(8) $\lambda_{ij}{}^{kl}$ and of complex forms $\sigma_{ijkl}$. 
The indices $i,j,\ldots$ denote the $\bf 8$ and $\bf\bar8$ representations. 
The change of basis is performed with the help of chiral $\Gamma$ matrices that interpolate between the two sets of indices, taking advantage of the triality property of the common subgroup SO$(8)$:
\begin{equation}\label{cayley matrix and complex basis}
 S_{\underline M}{}^N \equiv 
  \frac{1}{4\sqrt2}
  \begin{pmatrix}
     \Gamma_{ij}{}^{AB} &  i\Gamma_{ijAB} \\
  	 \Gamma^{ijAB}      & -i\Gamma^{ij}{}_{AB} \\
  \end{pmatrix}
,\quad
 [t_\alpha]_{\underline M}{}^{\underline N}
 \equiv
  \begin{pmatrix}
    \lambda_{ij}{}^{kl} & \sigma_{ijkl} \\
    \bar\sigma^{ijkl} & \bar\lambda^{ij}{}_{kl} \\
  \end{pmatrix}
 =
 S_{\underline M}{}^P [t_\alpha]_{P}{}^{Q} S^\dagger_Q{}^{\underline N}.
\end{equation}
From now on, underlined indices refer to the complex basis and allow us to keep track of
transformation properties with respect to (local) SU$(8)$ transformations. 
In order to reproduce the expected change of basis, one also has to fix the self-duality properties:
\begin{equation}
 [\Gamma_{ABCD}]_i{}^j = -[\star\Gamma_{ABCD}]_i{}^j,\quad
 [\Gamma_{ijkl}]_A{}^B = \eta[\star\Gamma_{ijkl}]_A{}^B,\ \eta=\pm1.
\end{equation}
The first anti-selfduality condition mirrors the anti-selfduality of SU(8) generators $\Sigma_{ABCD}$.
We refer to the second condition as $\eta$-selfduality.

The scalar manifold is a coset space and hence coset representatives can be employed to construct its explicit parameterization.
Using for E$_{7(7)}$ the complex basis we just introduced, E$_{7(7)}$/SU(8) representatives have the form:
\begin{equation}\label{coset representatives complex basis}
  L(\phi)_{\underline M}{}^{\underline N}
  = \exp\begin{pmatrix}
      0 & \phi_{ijkl} \\
      \phi^{ijkl} & 0 \\
      \end{pmatrix}.
\end{equation}
However, for future convenience we will choose a mixed basis to define the
representatives:
\begin{equation}\label{coset representatives mixed basis}
 L(\phi)_M{}^{\underline{N}} \equiv 
 S^\dagger_{M}{}^{\underline P} L(\phi)_{\underline{P}}{}^{\underline{N}}.
\end{equation}
In this way we can use coset representatives to switch between SL(8)  and
SU(8) covariant objects. This is the optimal setup to discuss gaugings of
subgroups of SL(8).
Actually, when discussing only the bosonic sector of the theory, it is often useful to rewrite the theory in terms of SU(8) invariant objects and hence describe the scalar manifold via the symmetric matrix
\begin{equation}
	{\cal M} = L L^T.
\end{equation}
This matrix is manifestly SU(8) invariant and transforms linearly under E$_{7(7)}$ transformations
\begin{equation}
	\delta {\cal M} = \Lambda \,{\cal M} + {\cal M}\, \Lambda^T,
\end{equation}
where $\Lambda = \Lambda^\alpha\, (t_{\alpha})_M{}^N$.

The two key ingredients for our analysis, the scalar kinetic term and the scalar potential, are expressed in terms of ${\cal M}$ as
\begin{equation}
	{\cal L}_{scalars} = \frac18\, {\rm Tr}(\partial_\mu {\cal M} \partial^\mu {\cal M}^{-1}) - \frac{g^2}{672}\left( X_{MN}{}^{R} X_{PQ}{}^{S} {\cal M}^{MP} {\cal M}^{NQ} {\cal M}_{RS}+ 7 \,X_{MN}{}^{Q} X_{PQ}{}^{N} {\cal M}^{MP} \right),
\end{equation}
where $X_{MN}{}^P$ define the structure constants of the gauge algebra.
As expected, the scalar potential crucially depends on the choice of the gauge group and in particular on its generators, which form a subalgebra of the algebra of duality transformations.
Actually, almost all the information needed to construct the full $N=8$ lagrangian can be encoded in a single tensor $\Theta_M{}^\alpha$ that specifies which generators of the duality algebra have been chosen as generators of the gauge algebra (or how the gauge algebra is embedded in the duality algebra, hence the name embedding tensor):
\begin{equation}
  X_{M} = \Theta_{M}{}^{\alpha} t_{\alpha}.
\end{equation}
Since $\Theta$ determines $X$, it also determines the scalar potential.

Obviously, the choice of $\Theta$ is constrained by consistency requirements and by supersymmetry.
The first obvious constraint has to do with the fact that the theory has at most 28 vector fields and therefore the dimension of the gauge group, given by the rank of $\Theta_M{}^\alpha$ cannot exceed 28. Moreover, we would like the theory to be local.
This is translated into the quadratic constraint
\begin{equation}
	\label{quad1}
  \Theta_{M}{}^{\alpha}\Theta_{N}{}^{\beta} \Omega^{MN} = 0,
\end{equation}
which is telling us which 28 linear combinations of the 56 electric and magnetic vector fields are used for the gauging procedure and hence which of the 56 linear combinations in $X_M$ are independent.
The second set of constraints comes from the request that the $X_M$ generators close into an algebra
\begin{equation}\label{quad2_closure}
	\left[X_M, X_N\right] = -X_{MN}{}^P X_P,
\end{equation}
which is proved to be equivalent to the request for the embedding tensor $\Theta$ to be invariant under gauge transformations, namely the quadratic constraint
\begin{equation}
	\label{quad2}
	X_{PM}{}^N \Theta_N{}^\alpha + \Theta_P{}^\beta  \Theta_M{}^\gamma \,f_{\beta \gamma}{}^\alpha =0,
\end{equation}
where $f_{\alpha \beta}{}^\gamma$ are the structure constants related to the adjoint representation of the generators $(t_{\alpha})_{\beta}{}^\gamma = - f_{\alpha \beta}{}^\gamma$.
The $X_{MN}{}^P$ are the gauge generators $X_M$ in the $\bf56$ representation. In \eqref{quad2_closure} they also play the role of the gauge structure constants, however we stress that they may also contain a symmetric part $X_{(MN)}{}^P\neq0$, which vanishes upon contraction with $X_P$. 
They also generically fail to obey the Jacobi identity, by a term which again vanishes upon contraction with the generators.

Finally, supersymmetry imposes a set of linear constraints:
\begin{equation}
	\label{lin}
  t_{\alpha M}{}^{N} \Theta_{N}{}^{\alpha} = 0, \qquad (t_{\beta}t^{\alpha})_{M}{}^{N}\Theta_{N}{}^{\beta} = - \frac12 \, \Theta_{M}{}^{\alpha},
\end{equation}
where the index $\alpha$ has been raised with the inverse group metric.

Altogether (\ref{quad1}), (\ref{quad2}) and (\ref{lin}) constrain $\Theta$, so that the allowed embedding tensors will be matrices with rank less or equal to $28$ that are defined by parameters in the $\mathbf{912}$ of E$_{7(7)}$, according to the decomposition
\begin{equation}
	\mathbf{56} \times \mathbf{133} = \mathbf{56} + \mathbf{912} + \mathbf{6480}.
\end{equation}
One can also prove that once the linear constraint is satisfied and the embedding tensor sits in the $\bf912$ representation, the two quadratic constraints (\ref{quad1}) and (\ref{quad2}) are equivalent.

\subsection{Supersymmetry breaking patterns} % (fold)
\label{sub:supersymmetry_breaking_patterns}

Before embarking ourselves in the task of extremizing the scalar potential scanning among all possible consistent solutions of the previous constraints on the embedding tensor, we would like to pause and see what are the possible supersymmetry breaking patterns one can expect.

A generic extremum of the scalar potential is going to break partially or completely supersymmetry, once more according to the entries of the embedding tensor, which also define the fermion shifts of the supersymmetry transformation rules.
In fact these are given by contractions of the so-called T-tensor, which can be defined in terms of the embedding tensor and the coset representatives as
\begin{equation}
	\label{Ttensor}
	T_{\underline{MN}}{}^{\underline P} = L^{-1}_{\underline M}{}^QL^{-1}_{\underline N}{}^R \Theta_Q{}^\alpha (t_{\alpha})_R{}^S L_S{}^{\underline P}.
\end{equation}
In the SU(8) basis mentioned above the content of the T-tensor can be expressed by two tensors $A_{1,2}$ as
\begin{equation}
T_i{}^{jkl} = L^{-1\,kl\,M} L^{-1}_{mi}{}^{N} X_{MN}{}^P L_P{}^{mj}
= -\frac{3}{4}A_2{}_i{}^{jkl} +\frac{3}{2} \delta_i^{[k} A_1^{l]j},
\end{equation} 
so that the supersymmetry transformation rules of the fermions become 
\begin{subequations}
		\label{susyrel}
	\begin{align}
	\delta\psi_\mu^{i} &= 2D_\mu \epsilon^{i}+\ldots+ \sqrt2 g A_1^{ij} \gamma_\mu \epsilon_j,\\[3mm]
	\delta \chi^{ijk} &= \ldots - 2 g A_{2\,l}{}^{ijk} \epsilon^l .
\end{align}
\end{subequations}
We will explicitly check how many residual supersymmetries are left on the vacua we will find in our analysis.
However, it is useful to know in advance which supersymmetry breaking patterns we can expect from a general analysis of the multiplet structure.
In particular, while we know that it is not difficult to break $N=8$ supergravity to an $N = 1$ or $N=3$ supergravity theory on an anti-de Sitter spacetime, there are no examples that we know of where supersymmetry is broken to an odd number preserving a vanishing cosmological constant.
Following \cite{Andrianopoli:2002rm,Andrianopoli:2002vy} one can regroup the matter content of the $N=8$ supergravity multiplet into smaller multiplets of $N' < N$ supersymmetry.
One obvious constraint is given by the fact that there is always one gravity multiplet (with $N'$ gravitini) and that there are always $8-N'$ massive gravitino multiplets.
If, after completion of the above multiplets, there are still matter fields left, one should try to accommodate them in short and long representations of the residual supersymmetry group, also taking into account that the Goldstone bosons of the broken gauge symmetry are related to the difference of translational symmetries between the $N=8$ and the $N =N'$ scalar manifold \cite{Andrianopoli:2002rm,Andrianopoli:2002vy}.

Performing such an analysis it is almost straightforward to show that, for zero vacuum energy, while partial supersymmetry breaking to an even number of supersymmetries is kinematically possible, partial supersymmetry breaking to an odd number of supersymmetries is extremely constrained.
Vacua with $N=5,7$ are forbidden \cite{Andrianopoli:2002rm,Andrianopoli:2002vy} and there is only one possible pattern of supersymmetry breaking to $N=3$.
In order to break $N=8 \to N' = 3$ one has to regroup the fields of the $N=8$ supergravity multiplet into $N'=3$ multiplets.
The $N=8$ supergravity multiplet can be represented as
\begin{equation}
	\{ (2), 8 \cdot (3/2), 28 \cdot (1), 56 \cdot (1/2), 70 \cdot (0)\},
\end{equation}
where the notation $n \cdot (s)$ means that there are $n$ fields of spin $s$.
The relevant $N'=3$ multiplets on the other hand are:
\begin{eqnarray}
 {\rm gravity}: && \{(2), 3 \cdot (3/2), 3 \cdot (1), (1/2) \}	, \\[2mm]
 {\rm long\ gravitino}: && \{(3/2), 6 \cdot (1), 14 \cdot (1/2), 14\cdot(0) \}	, \\[2mm]
 {\rm semilong\ gravitino}: && 2 \times \{(3/2), 4 \cdot (1), 6 \cdot (1/2), 4\cdot(0) \}	, \\[2mm]
 {\rm massless\ vector}: && \{(1), 4 \cdot (1/2), 6 \cdot (0) \}	, \\[2mm]
 {\rm massive\ vector}: && 2 \times \{(1), 4 \cdot (1/2), 5 \cdot (0) \}.
\end{eqnarray}
From this structure we deduce that there are only two options for decomposing the $N=8$ supergravity multiplet into $N'=3$ multiplets, either 1 graviton, 4 semilong gravitini, 1 long gravitino and 3 massless vector, or 1 graviton, 4 semilong gravitini, 1 long gravitino, 1 massless vector and 2 massive vectors.
However, once massive states have been integrated out the remaining scalar manifold would be 
SU(3,3)/[SU(3)$\times$SU(3)$\times$U(1)] in the first case and SU(3,1)/[SU(3)$\times$ U(1)] in the second case.
The constraint on the translational symmetries shows that only the second option is consistent.
Indeed, in the first case one has given mass to 22 vector fields, but only 18 flat directions have been integrated out.
On the other hand, in the second case one has given mass to 24 vector fields and got rid of 24 out of the 27 original translational isometries of the scalar manifold.
From this we expect that there is a unique way of breaking $N=8$ supersymmetry to $N'=3$ preserving a Minkowski vacuum.

Different patterns are allowed if one investigates $N=8 \to N=1$.

% subsection supersymmetry_breaking_patterns (end)

% section n_8_supergravity (end)

\section{Finding extrema of the scalar potential} % (fold)
\label{sec:finding_extrema_of_the_scalar_potential}

We now come to the main part of this note, where we are going to discuss the procedure we are going to adopt to find extrema of the scalar potential
\begin{equation}\label{scalar potential}
	V(\phi) = \frac{g^2}{672}\left( X_{MN}{}^{R} X_{PQ}{}^{S} {\cal M}^{MP} {\cal M}^{NQ} {\cal M}_{RS}+ 7 \,X_{MN}{}^{Q} X_{PQ}{}^{N} {\cal M}^{MP} \right).
\end{equation}
This potential is an obvious function of the 70 scalar fields $\phi$, via the coset representatives $L$ (and the SU(8) invariant matrix ${\cal M}$), but also of the embedding tensor $\Theta_M{}^\alpha$, which defines the structure constants $X_{MN}{}^P$.
Explicitly
\begin{equation}
	V(\phi) = V(L(\phi), \Theta).
\end{equation}
Vacua of $N=8$ supergravity can be obtained by finding extrema of $V(\phi)$ with respect to $\phi$, i.e.~by solving a coupled system of 70 algebraic equations
\begin{equation}
	\frac{\partial V(\phi)}{\partial \phi^{ijkl}} = 0.
\end{equation}
For a generic case, the dependence of the scalar potential on the scalar fields is rather complicated.
In the best case, where we imagine to employ the Iwasawa decomposition to perform a solvable parameterization of the scalar manifold, $V(\phi)$ contains a combination of exponential functions of the scalar fields associated to the Cartan generators and polynomial functions of the scalars associated to the nilpotent generators.
The obvious consequence is that finding solutions to such a complicated system of equations is challenging and one usually restricts one's attention to subsectors of the scalars, invariant under specific symmetry groups, in order to simplify the task.
Also in this case solutions can often be provided only numerically.

However, there is an alternative that simplifies the computation of the extrema of the scalar potential and maps the system of scalar equations to a coupled set of second and first order conditions on the gauging parameters.
This alternative \cite{Inverso,Dibitetto:2011gm} relies on the simple observation that the scalar manifold is a coset space and that therefore each point on the manifold can be mapped to any other by an E$_{7(7)}$ transformation and, most importantly, on the observation that the scalar potential is invariant under a combined action of these transformations on the coset representatives $L$ and on the embedding tensor $\Theta$.
In detail, the scalar potential depends only on the contracted combination\footnote{Here $L^{-1} \Theta$ formally expresses contractions on both indices of the embedding tensor.} $L^{-1}\Theta$:
\begin{equation}
	V(\phi) = V(L^{-1} \Theta).
\end{equation}
Since we are allowed to map any point of the manifold into any other via a duality transformation, for a given scalar potential we can map any critical point to the ``origin'', namely the $\phi =0$ point.
At such point, the scalar potential is a simple quadratic function of the structure constants and the minimization conditions become quadratic conditions on the embedding tensor $\Theta$ that should be solved together with the set of conditions (\ref{quad1}), (\ref{quad2}) and (\ref{lin}) defining consistent gaugings.
This means that rather than fixing the gauging and then performing a scan of all possible critical points of the scalar potential and then scan among the possible gaugings, one can simply solve a set of quadratic conditions on the embedding tensor and then read the resulting values of $\Theta$ which define at the same time the original gauge group, the value of the cosmological constant and the masses at the critical point.

% \begin{wrapfigure}{r}{0.55\textwidth} 
\begin{figure}[htb]
% \vspace{-9mm}
\begin{center}
	\includegraphics[scale=.55]{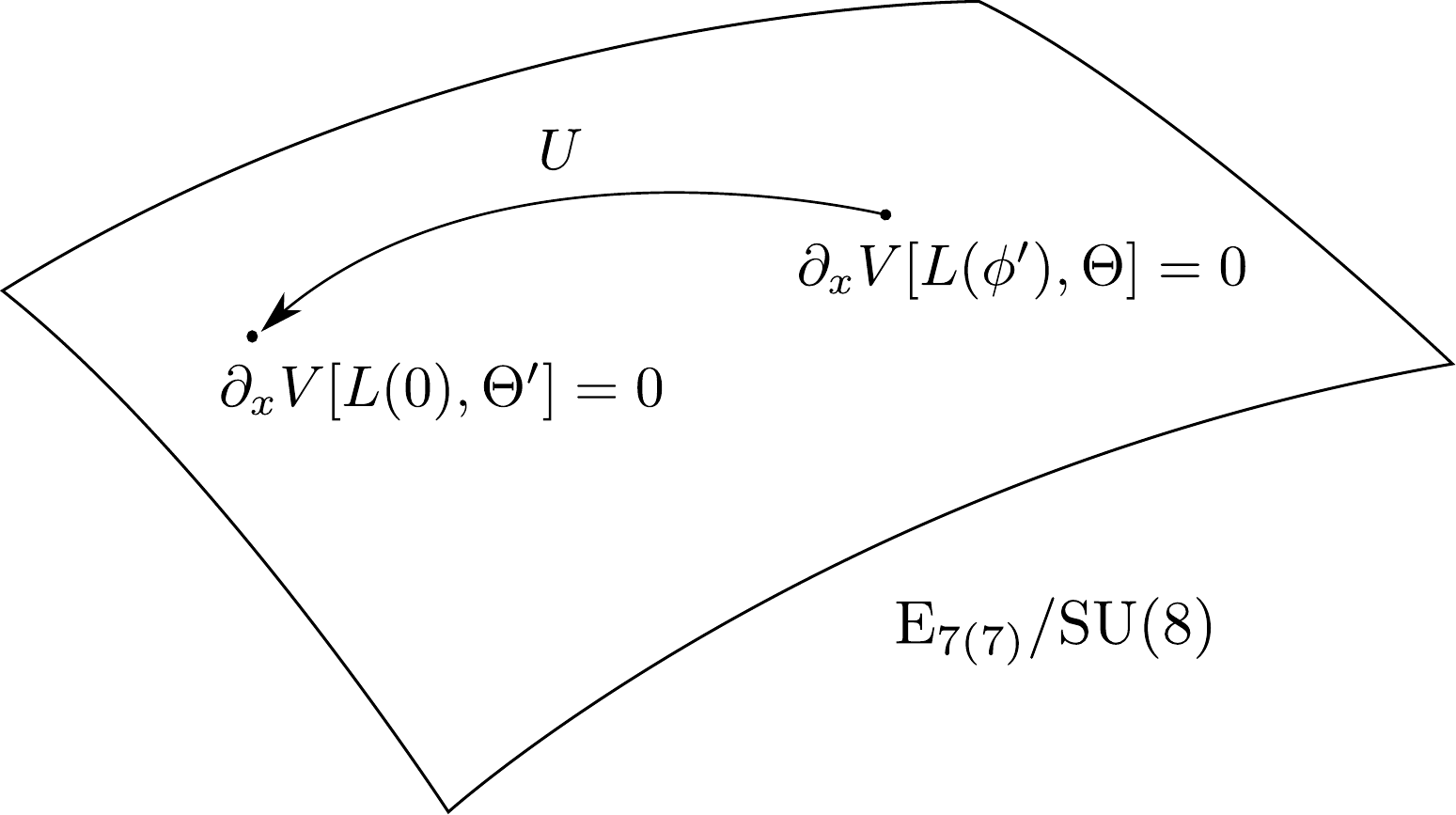} 
\end{center}
% \vspace{-7mm}
\caption{\small \emph{Stationary points of the scalar potential found for $\phi = \phi'$ can be translated to the fixed point of the scalar manifold $\phi = 0$ by a non-compact isometry transformation $U$ and by a redefinition of the embedding tensor $\Theta' = U \Theta$.}} \label{potplot} 
\end{figure}
% \end{wrapfigure}
As mentioned above, in order for this procedure to work, one should be able to redefine the scalar potential so that its value at any point is mapped to the origin. 
This happens if one considers a combined action of the E$_{7(7)}$ duality group on the coset representative and on the embedding tensor.

The action of an element $U \in $ E$_{7(7)}$ on the scalar fields is linear on the coset representatives
\begin{equation}
	U L(\phi) = L(\phi') h(\phi, \phi'), \quad {\rm with } \; h\in {\rm SU(8)}.
\end{equation}
Since the scalar potential depends only on the SU(8) invariant matrices ${\cal M}$, we see that the extra field-dependent SU(8) transformation $h$ leaves $V(\phi)$ invariant.
If we act with the same generator also on the embedding tensor
\begin{equation}
	\Theta \to \Theta' = U \Theta,
\end{equation}
we see that also the scalar potential remains invariant.
On the other hand we can interpret an action by $U \in $ E$_{7(7)}$ on the embedding tensor as an isometry transformation on the scalar fields.
In detail, since
\begin{equation}\label{iso transformation acting on L and Theta}
	L^{-1}(\phi) \Theta' = L^{-1}(\phi) U \Theta = h L^{-1}(\phi') \Theta,
\end{equation}
we get that the scalar potential in terms of the transformed embedding tensor is equivalent to the scalar potential evaluated at a different point on the scalar manifold:
\begin{equation}
	V(L(\phi), \Theta') = V(L(\phi'), \Theta).
\end{equation}
This identity allows us to calculate the scalar potential and its derivatives for one single value of the scalar fields, for instance $\phi = 0$, and the embedding tensor in the form that defines a particular gauge group. 
Then we modify the embedding tensor in order to recover information on the rest of the scalar manifold. 
The fact that we can actually cover the whole manifold with this strategy is guaranteed by the property of coset manifolds of being homogeneous spaces. 
The choice of $\phi = 0$ as the point in which we perform all the calculations has the advantage that it is a fixed point under the action of the maximal compact subgroup of isometry transformations SU(8), so that we can consider modifications of $\Theta$ related only by the non-compact transformation and take the constant transformation $U$ to belong itself to the coset space E$_{7(7)}$/SU(8), so that there is a one-to-one correspondence between transformations $U$ and independent directions on the scalar manifold.

More precisely, the scalar potential is defined starting from the irreducible parts of the T-tensor (\ref{Ttensor}) and different embeddings of the same gauge group can be related by E$_{7(7)}$ transformations $U$, so that, at the level of the gauge algebra generators
\begin{equation}
	X_{MN}{}^P  = \Theta_M{}^\alpha (t_{\alpha})_N{}^P \  \to\  X'_{MN}{}^P = U_M{}^Q U_N{}^R X_{QR}{}^S U^{-1}{}_S{}^P.
\end{equation}
As stated previously, we can see this as a redefinition of the embedding of the gauge group or as a change of coordinates on the scalar manifold, so that the T-tensor gets redefined
\begin{equation}\label{T-tensor redefinition}
	T'{}_{\underline{MN}}{}^{\underline P}(0) = T_{\underline{MN}}{}^{\underline P}(\phi_U), 
\end{equation}
where $U^{-1} L(0) = L(\phi_U) h(\phi_U)$ and where $T'$ refers to the T-tensor computed from the $X'$ structure constants, while $T$ is the T-tensor computed from $X$.
The equivalence \eqref{T-tensor redefinition} actually holds up to the action of $h(\phi_U)$ on the right-hand side, which can either be eliminated by a suitable choice of $U$ or by noting that SU(8) is a symmetry of the action and in particular that the scalar potential is SU(8) invariant.
Furthermore we can see that, up to similarity transformations on the E$_{7(7)}$ generators $t_\alpha \to U t_\alpha U^{-1}$, the only action we really need to take into account is the left action on $\Theta_M{}^\alpha \to  \Theta'{}_M{}^\alpha = U_M{}^N \Theta_N{}^\alpha$.
Hence we see once more that changing the embedding of $G_g$ inside E$_{7(7)}$ can be taken to correspond to keeping the old embedding and moving on the scalar manifold. 
For our purposes this means that we can investigate the presence of vacua for any value of the scalar fields just by sitting at the origin of the coset manifold and varying the embedding of the gauge group.

We can apply the same principle to explicitly compute the derivatives of $V(\phi)$ at the base-point of the scalar manifold $\phi=0$.
There we can take the vielbeins to be proportional to the identity matrix so that the derivatives of the coset representatives with respect to the 70 scalar fields $\phi^{ijkl}$ become proportional to the 70 non-compact generators of E$_{7(7)}$, $t_{ijkl}$:
\begin{equation}
	\partial_{ijkl}L(0)_{M}{}^N \propto [t_{ijkl}]_M{}^P L_P{}^N(0).
\end{equation}
We can then use this information to simplify the computations of the derivatives of the scalar potential.
The coset representatives appear in the scalar potential always contracted with the embedding tensor and therefore  the derivative of the contracted expression at the base-point can be computed by an appropriate action of the non-compact generators.
Formally
\begin{equation}
	\partial_{ijkl}L^{-1}(0)\Theta \propto L^{-1}(0)t_{ijkl}\Theta,
\end{equation}
where the action of the generators should be taken on all the indices of the embedding tensor, as both of them are contracted (via the generators) to the coset representatives.
It is therefore straightforward to replace the derivatives of the scalar potential with respect to the scalar fields with the derivatives with respect to the embedding tensor, replacing the embedding tensor by the product of the non-compact generators with the embedding tensor itself:
\begin{equation}
		\partial_{\gamma} V(0,\Theta) \propto \frac{\delta V}{\delta \Theta} \cdot \left[ (t_{\gamma})_M{}^M 
		\Theta_N{}^\alpha + \Theta_M{}^\beta f_{\gamma \beta}{}^\alpha \right].
\end{equation}
Here we replaced the multi-index $ijkl$ with $\gamma$, an index that runs only on the non-compact $E_{7(7)}$ generators, while $\alpha$ and $ \beta$ run over all the $E_{7(7)}$ generators.
If, in addition, we choose an orthonormal basis for the ${\mathfrak e}_{7(7)}$ generators and a frame where ${\cal M}(0)=\mathbb1$, we can simplify the explicit expressions for the potential and its derivatives.
The scalar potential (\ref{scalar potential}) at the origin reduces to
\begin{equation}
	\label{V0}
	V(0,\Theta) = \Theta_{M}{}^\alpha \Theta_M{}^\beta\, (\delta_\alpha^\beta + 7\, \eta_{\alpha\beta}),
\end{equation}
where $\eta_{\alpha\beta}$ is the Cartan--Killing metric on ${\mathfrak e}_{7(7)}$.
Note that, as expected, the expression is not E$_{7(7)}$ invariant. 
The first derivative with respect to the scalar fields at $\phi = 0$ is proportional to
\begin{equation}
	\label{first derivative as isometry variation}
	\partial_\rho V(0,\Theta) \quad \propto \quad \Theta_{M}{}^\alpha [t_\rho]_{M}{}^{N}\Theta_N{}^\beta(\delta_\alpha^\beta + 7 \eta_{\alpha\beta})
			+\Theta_{M}{}^\alpha \Theta_M{}^\beta f_{\rho\alpha}{}^{\beta},
\end{equation}
where $\rho$ runs only over the 70 non-compact generators. 
Finally, the mass matrix can also be simplified to the following expression
\begin{equation}
	\label{mass matrix as isometry variation}
\begin{split}
	M_\rho{}^{\chi}(0,\Theta) \propto&\ \Theta_{M}{}^\alpha (t_\rho t^\chi)_M{}^N\Theta_N{}^\beta(\delta_\alpha^\beta + 7 \eta_{\alpha\beta}) +\Theta_{M}{}^\alpha\Theta_M{}^\beta (f_{\rho}f^{\chi})_\beta{}^{\alpha} \\[2mm]
&+\Theta_{M}{}^\alpha [t_\rho]_{M}{}^{N} \Theta_N{}^\beta f^\chi{}_{\beta}{}^{\alpha}
+\Theta_{M}{}^\alpha [t^\chi]_{M}{}^{N} \Theta_N{}^\beta f_{\rho\beta}{}^{\alpha} .
\end{split}
\end{equation}
The proportionality factor can be fixed by the analysis of the scalar kinetic term, as we will see later.

We would also like to identify the residual gauge symmetry at the vacua.
Given $g\in G_g$ and being $\phi$ an extremal point of $V$, the usual condition $g(\phi)=\phi$ is best written in terms of the representatives:
\begin{equation}\label{residual sym condition at phi}
 g L(\phi) = L(\phi)h.
\end{equation}
According to \eqref{iso transformation acting on L and Theta}, we want to move the stationary point back to $\phi'=0$ with a global transformation $U$ acting on both $L(\phi)$ and $\Theta$, so that
\begin{equation}
 L(\phi)\rightarrow L(0)h=UL(\phi),\quad \Theta\rightarrow\Theta'= U\Theta.
\end{equation}
We remind that also the embedding of the gauge group inside E$_{7(7)}$ is affected by $U$, so that the gauge group defined by $\Theta'$ is isomorphic to the one given by $\Theta$, namely $G'_g= UG_gU^{-1}$.
We can then rewrite \eqref{residual sym condition at phi} in terms of $g'\in G'_{g}$:
\begin{equation}
 g L(\phi) = U^{-1}g' L(0) h
\end{equation}
Since $L(\phi)=U^{-1}L(0)h^{-1}$, the residual symmetry condition equivalent to  \eqref{residual sym condition at phi} reads:
\begin{equation}\label{residual sym condition at 0}
 g' L(0) = L(0) h \quad \Leftrightarrow \quad g'\in SU(8)
\end{equation}
We see that the residual symmetry is given by the intersection of the gauge group $G'_{g}$ generated by $X'_{MN}{}^P$ with SU$(8)$.

The expression of the scalar potential at the origin of the moduli space (\ref{V0}) triggers a simple observation on the allowed values of the cosmological constant on the vacuum.
Since the Cartan--Killing metric is negative definite on the compact generators and it is non-trivial for semisimple groups, one obtains immediate consequences on the expected types of vacua.
For instance, in the case of SO($p,q$) gaugings and for canonically normalized generators, the cosmological constant is proportional to the weighted difference between the compact and the non-compact generators: $V \sim 8\cdot (p q) - 6\cdot (p(p-1)+q(q-1))/2$.
This means that we should expect de Sitter vacua only for SO(5,3) and SO(4,4) gaugings and Minkowski vacua only for SO(6,2).
However, this computation is valid only in a basis where the generators are canonically normalized at the origin. Other vacua may appear at different points of the moduli space and once we move them at the origin the normalizations of the various generators change (as dictated by the new values of the embedding matrices).
In the original picture, this means that moving sufficiently far away from the origin one can change the sign of the potential.
Although this implies that we do not have a conclusive a priori argument for the value of the cosmological constant, this simple observation explains why most of the vacua found so far respect this pattern (see Tables~\ref{tab_eigenvalue solutions only theta} and \ref{tab_riassunto} summarizing our results for such groups).

Finally, let us recall on a simpler and more restrictive scenario noted in \cite{Dibitetto:2011gm}.
Whenever the non-vanishing terms of the embedding tensor are mapped among themselves under duality transformations, one can consider the analysis at the origin of the moduli space as exhaustive of all the vacua for that type of gauging.

% section finding_extrema_of_the_scalar_potential (end)

\subsection{Symplectic frames} % (fold)
\label{sec:symplectic_frames}

As explained previously, the gauging procedure is specified by the embedding tensor $\Theta$.
This tells us which linear combinations of the vector fields realize the local gauge group.  
In 4 dimensions, however, vector fields are dual to themselves.
This means that in 4-dimensional supergravity the same gauge group could be gauged by minimal electric couplings to the 28 vector fields in the lagrangian before gauging or also by (some of) the dual magnetic fields.
This means that the choice of symplectic frame (i.e.~which vectors are electric and which are magnetic) provides important information on the description of the gauging and therefore also on the vacua of the theory. 
This is the only extra freedom that one has in specifying a maximal supergravity model and is described in terms of representatives of GL(28)$\backslash$Sp(56,$\mathbb R$)$/$E$_{7(7)}$, which defines the allowed embeddings of the duality group in Sp(56,$\mathbb R$) \cite{deWit:2007mt}.
The split of the vectors into electric and magnetic ones is related to the splitting of the fundamental representation of E$_{7(7)}$, namely  the one of dimension $\mathbf{56}$, into two distinct sets of 28 vector fields: $A_\mu^M = (A_\mu^{\Lambda},  A_{\mu\,\Lambda})$, where $A_\mu^\Lambda$ transform in some definite representation of the group $G_0$ of global symmetries of the lagrangian.
For any choice of the gauge group $G_g$, there is at least one electromagnetic frame in which this (modulo abelian ideals) is realized as a subgroup of $G_0$.
This is the electric frame, where we can consistently set $\Theta^{\Lambda \alpha} = 0$ and the quadratic constraint (\ref{quad1}) is automatically satisfied.

An important observation is that the action of the duality group on the embedding tensor (which is crucial in our minimization procedure described above) may in general turn on couplings to magnetic vector fields $A_{\mu\,\Lambda}$ even if they were initially absent.
For instance, if we had considered an electric gauging of some group and hence set $\Theta^{\Lambda \alpha} = 0$, a generic action on $\Theta$ by E$_{7(7)}$ would introduce non-vanishing magnetic couplings
\begin{equation}
	\Theta'_M{}^{\alpha} = U_M{}^N \Theta_N{}^\alpha = \left(\begin{array}{cc}
	A & B \\ C & D
	\end{array}\right) \left(\begin{array}{c}
	\Theta_{\Lambda}{}^\alpha \\ 0
	\end{array}\right)
\end{equation}
unless $C=0$, which restricts $U$ to the group of global invariances of the original lagrangian $G_0$.
For this reason, imposing  $\Theta^{\Lambda \alpha} = 0$ or any other specific choice would generally be a too strong restriction on our analysis.
On the other hand, since we are not interested to provide already in this paper a complete classification of all possible gaugings and their vacua, we decided to work allowing also magnetic vectors to enter the gauge connection, but we restricted the number of entries in $\Theta_M{}^\alpha$ that are allowed to be turned on, as we we will now explain in detail.

We stress that although a change in electromagnetic frame does not represent a symmetry of the theory (not even at the level of the equations of motion), it does not affect the scalar potential.
Therefore fixing a choice of electromagnetic frame does not in principle reduce the generality of the analysis and has the only effect of allowing $\Theta_M{}^\alpha$ to be easily decomposed with respect to representations of a chosen electric group $G_0$.

A natural choice for the symplectic frame is one that allows for a direct comparison of our results with the already known ones, in particular those related to the vacua of the CSO($p,q,r$) gaugings.
For this reason we used the so-called SL(8,${\mathbb R}$) basis, where the E$_{7(7)}$ generators are decomposed according to the embedding of one of its SL(8,${\mathbb R}$) subgroups, so that the $\mathbf{56}$ electric and magnetic vector fields split into $\mathbf{28}$ electric and $\mathbf{28}'$ magnetic fields, according to the decomposition (\ref{sl8 decomposition}).
In detail the electric $A_\mu^{AB}$ and magnetic $A_{\mu \, AB}$ potentials 
% (where $A,B = 1,\ldots,8$ is an sl(8,${\mathbb R}$) index and $A_\mu^{AB} = - A_\mu^{BA}$) 
transform as
\begin{eqnarray}
	\delta A_\mu{}^{AB} &=& A_\mu{}^{CD} \Lambda_{CD}{}^{AB} + A_{\mu CD} \, \star\Sigma^{CDAB}, \\
	\delta A_{\mu\,AB} &=& A_{\mu\,CD} \Lambda^{\prime CD}{}_{AB} + A_\mu{}^{CD}  \, \Sigma_{CDAB},
\end{eqnarray}
where the action of the E$_{7(7)}$ generators follows from the decomposition (\ref{e7 gen in sl8 basis}) and the $\Lambda$ and $\Lambda'$ matrices are related to the $\mathfrak{sl}$(8,${\mathbb R}$) ones $\Lambda_A{}^B$, as explained there.

In the SL(8,${\mathbb R}$) frame, the embedding tensor components, given by the tensor product of the $\mathbf{56}$ and $\mathbf{133}$ representations of E$_{7(7)}$, are decomposed as 

\bigskip

\begin{center}
\begin{tabular}{c|cc}
& \textbf{28} & \textbf{28}$'$ \\[1mm]\hline
\textbf{63} & \textbf{36 + 420} & \textbf{36}$'$ + \textbf{420}$'$ \\[1mm]
\textbf{70} & \textbf{420}$'$ & \textbf{420} 
\end{tabular}
\end{center}

\bigskip

\noindent
The entries in the table actually denote representations conjugate to the ones in which the embedding tensor transforms. 
Since only one \textbf{420} representation appears in the branching\footnote{Under SL(8) we have the branching rule $\mathbf{912} \to \mathbf{36} + \mathbf{36}^\prime + \mathbf{420} + \mathbf{420}^\prime$.} of the \textbf{912}, the two \textbf{420} representations in the table must coincide, and so must the $\mathbf{420}'$ representations. 
This implies that, if $\Theta_M{}^\alpha$ had a contribution in the \textbf{420}, it would describe a coupling of the gauge fields to the generators in the \textbf{70}, but also, at the same time, induce a coupling of the dual gauge fields to the generators in the \textbf{63} of SL(8,${\mathbb R}$).

In the case of purely electric gaugings, this analysis restricts the allowed embedding tensor to just the $\Theta_{AB}{}^C{}_D$ components.
Actually, electric gaugings are described by a symmetric tensor in the $\mathbf{36}'$ representation of SL(8,${\mathbb R}$), which is called $\theta_{AB}$ \cite{deWit:2002vt}.
Also, the embedding tensor is completely specified by 
\begin{equation}
	\label{sl8embedding}
	\Theta_{AB}{}^C{}_D = \delta_{[A}^C \, \theta^{\phantom{A}}_{B]D},
\end{equation}
while all the other components are set to zero.
The quadratic constraint (\ref{quad1}) is obviously identically satisfied for such gaugings, hence any symmetric real matrix $\theta_{AB}$ defines a consistent gauging, even in the case of a non-invertible matrix.
While all entries could be non-vanishing, the gauge groups resulting from $\theta \neq 0$ depend only on the number of positive ($p$), negative ($q$) and zero ($r$) eigenvalues.
In fact the electric gaugings in such a frame are restricted to the SO($p,q$) groups when $r=0$, defined as the finite SL(8) transformations $g$ that leave $\theta$ invariant: $g \theta g^T = \theta$. 
When $r\neq0$ one obtains the group contractions CSO($p,q,r$) \cite{hep-th/9804056,Hull:2002cv}. 
The actual gauge group in this case is $SO(p,q)\ltimes T^{r(8-r)}$.
The total number of inequivalent gaugings is 24, which is calculated taking into account that the gauge group does not change if we just permute the eigenvalues between them, nor if we change an overall sign (which would correspond to exchanging $p$ and $q$). 
The permutations modify the embedding of the gauge group in E$_{7(7)}$, but in a completely trivial way that does not affect in any way our discussion of the stationary points.

In our analysis we will also consider gaugings where the $\mathbf{36}$ representation is turned on.
We will describe the corresponding gauging parameters by the symmetric tensor $\xi^{AB}$.
Following once more the decomposition of the embedding tensor representations presented above, these parameters turn on the magnetic components of the embedding tensor:
\begin{equation}
	\label{sl8 embedding of xi}
	\Theta^{AB\,C}{}_D = \delta^{[A}_D \, \xi_{\phantom{A}}^{B]C}.
\end{equation}
Choosing $\xi \neq 0$ while $\theta = 0$ is just a relabeling of the vector fields and brings us back to the previous discussion.
However, when both $\theta$ and $\xi$ are turned on, the quadratic constraint is not identically satisfied anymore and one gets new and interesting situations.
The quadratic constraint gives the relation
\begin{equation}
	\label{thetaxi}
	\delta_E^D \,\xi^{AB}\theta_{BC}  = \xi^{DB} \theta_{BE}\, \delta^A_C,
\end{equation}
from which also $\xi^{AB} \theta_{BC} = \frac18\, {\rm Tr}(\theta \xi) \, \delta^A_C$ follows.
From this we see that we have only two options.
Whenever the matrix $\theta$ is invertible, the constraint is solved by
\begin{equation}
	\xi = c\, \theta^{-1}, \quad c \in \mathbb R, \quad {\rm for} \quad {\rm det}\, \theta \neq 0
\end{equation}
On the other hand, if $\theta$ is not invertible it has a non-trivial kernel.
In this case the setup solving (\ref{thetaxi}) is given by taking $\xi$ non-zero only in the subspace defined by the kernel of $\theta$.
We will come back to this discussion later on when we analyze the vacua of such gaugings.

% subsection symplectic_frames (end)

\section{Vacua of the gauged theory} % (fold)
\label{sec:vacua_of_the_gauged_theory}

Now that we have described all necessary ingredients for our analysis, we can present some results obtained by using our procedure.

\subsection{$\theta \neq 0$} % (fold)
\label{sub:_theta_neq_0_}

The first scenario we considered is the simple instance where the gauge group is contained in the SL(8,${\mathbb R}$) electric frame.
This means that we considered vacua of the CSO($p,q,r$) gaugings.
Given that we look for vacua sitting at the origin of the moduli space, the restriction of considering only electric gaugings (and hence $\Theta^{AB\, \alpha} = 0$) forbids us to obtain all the vacua that in the standard frame have a non-vanishing value of the axions as will be clear shortly.

From the embedding tensor defined in (\ref{sl8embedding}) we can easily recover the form of the structure constants  from the definition $X_{MN}{}^P \equiv \Theta_M{}^\alpha (t_{\alpha})_N{}^P$.
In the SL(8) basis they read
\begin{equation}
	X_{AB\,M}{}^N = \left(\begin{array}{cc}
	-f_{AB\, CD}{}^{EF} &  \\ &f_{AB\, EF}{}^{CD}
	\end{array}\right), \qquad X^{AB} = 0,
\end{equation} 
where $f_{AB\, CD}{}^{EF} = 2 \sqrt{2} \, \delta_{[A}^{[E} \, \theta^{\phantom{A}}_{B][C} \, \delta_{D]}^{F]}$, with the normalization chosen for later convenience, are the structure constants of CSO($p,q,r$).
The scalar potential and its first and second derivatives can then be computed by replacing the embedding tensor with its explicit form in (\ref{V0}), (\ref{first derivative as isometry variation}) and (\ref{mass matrix as isometry variation}).
For instance, the scalar potential, which fixes the value of the residual cosmological constant at the vacua, is
\begin{equation}
	\label{potential with only theta}
 V(\phi=0,\theta) = \frac{1}{4}\text{Tr}(\theta^{2}) -\frac{1}{8}(\text{Tr}\,\theta)^{2}.
\end{equation}
For the first and second derivative, however, we take here a different route which allows a better comparison with the literature, an indirect check of the formulae (\ref{first derivative as isometry variation}) and (\ref{mass matrix as isometry variation}) and a simple fix for the proportionality factors.

Following \cite{deWit:1983gs}, the scalar potential can be expressed in terms of the squares of the fermion shifts, which are defined in terms of contractions of the T-tensor.
They are in one to one correspondence with the decomposition $\bf912 \rightarrow 36+\overline{36}+420+\overline{420}$ of the T-tensor with respect to $SU(8)$ as shown before (\ref{susyrel}) and lead to the fermion shift tensors $A_{1,2}$.
At the origin of the moduli space and for the class of gaugings under consideration, these tensors are
\begin{equation}
\begin{split}
	A_{1}^{ij}(0) = \frac{\text{Tr}\theta}{8} \delta^{ij},\qquad
	A_{2\,i}{}^{jkl}(0)  = \frac{1}{8}\text{Tr}(\Gamma_{i}{}^{jkl}\theta).
\end{split}
\end{equation}
Coming back now to the derivatives of the scalar potential, one can see that the gradient of $V(\phi)$ is proportional to the $\eta$-selfdual part of a tensor $\Omega^{ijkl}\equiv\frac{3}{4}{A_{2\,m}}^{n[ij}{A_{2\,n}}^{kl]m}
	-{A_1}^{m[i} {A_{2\,m}}^{kln]}$, which in our current setup and for $\phi=0$ takes the form:
\begin{equation}\label{derivative as Omega tensor}
	\Omega^{ijkl}
%	\equiv\frac{3}{4}{A_{2\,m}}^{n[ij}{A_{2\,n}}^{kl]m}
%	-{A_1}^{m[i} {A_{2\,m}}^{kln]}
%\overset{\phi=0}{=}
     = \frac{1}{16}\text{Tr}(\Gamma^{ijkl}\theta^2)
            -\frac{1}{32}\text{Tr}(\theta)\text{Tr}(\Gamma^{ijkl}\theta).
\end{equation}
A similar expression exists for the second derivative which after some algebra reduces to
\begin{align}
	\label{hessian with only theta}
 6D_{ijkl}D^{mnpq}V\vert_{\phi=0}
 =&\
  \left(\frac{1}{8}\text{Tr}(\theta^2)+\frac{1}{16}(\text{Tr}\theta)^2\right)\delta_{ijkl}^{mnpq}
  -\frac{2}{3}A_{2}{}^{[m}{}_{[ijk}A_{2\,l]}{}^{npq]}\\\notag
 &+\frac{3}{16}\delta_{[ij}^{[mn}\text{Tr}\left(\Gamma^{pq]}\theta\Gamma_{kl]}\theta\right)
  -\frac{3}{4}\delta_{[ij}^{[mn}\text{Tr}\left(\Gamma_{k}{}^{p}\theta\Gamma^{q]}{}_{l]}\theta\right)
 -9\delta_{[ij}^{[mn}\Omega^{pq]}{}^{\phantom{[]}}_{kl]},
\end{align}
where the last term vanishes at the critical points of $V(\phi)$.
% We stress that we took two SU(8)-convariant derivatives instead of simple ones.
% Since $V(\phi)$ is SU(8) invariant, we can replace the first simple derivative by a covariant one because $V(\phi)$ is invariant.
% For the second derivative we used the fact that the extra terms are proportional to $V'$ and vanish at the stationary points.
The mass matrix of the scalar fields is then proportional to \eqref{hessian with only theta}.
However, in order to obtain the proper masses of the scalar fields, one should start from canonically normalized kinetic terms.
For the case at hand, the kinetic term of the scalar fields can be written as
\begin{equation}
-\left(\frac{1}{12}\cdot\frac{1}{24}\epsilon_{ijklmnpq}\right) \partial\phi^{ijkl}\partial\phi^{mnpq},
\end{equation}
where $\frac{1}{24}\epsilon_{ijklmnpq} $ plays the role of the metric with flat indices (we recall that $\phi^{ijkl}$ actually correspond to flat coordinates contracted with generators $\sigma^{ijkl}$).
It is then clear that the term $\left(\frac{1}{12}\cdot\frac{1}{24}\epsilon_{ijklmnpq}\right)$ plays the role of the kinetic matrix, ad its inverse should be contracted with $\frac{1}{2}\partial_{ijkl}\partial_{mnpq}V(0)$ to give the mass matrix $M_{ijkl}{}^{mnpq}$.
The final expression is
%\hspace{2em}
\begin{equation}
\begin{split}
M_{ijkl}{}^{mnpq}
\equiv&\
  \left(\frac{1}{8}\text{Tr}(\theta^2)+\frac{1}{16}(\text{Tr}\theta)^2\right)\delta_{\,i\,j\,k\,l}^{mnpq}
  -\frac{2}{3}A_{2}{}^{[m}{}_{[ijk}A_{2\,l]}{}^{npq]}\\[.2em]
 &+\frac{3}{16}\delta_{[ij}{}^{[mn}\text{Tr}\left(\Gamma^{pq]}\theta\Gamma_{kl]}\theta\right)
  -\frac{3}{4}\delta_{[ij}{}^{[mn}\text{Tr}\left(\Gamma_{k}{}^{p}\theta\Gamma^{q]}{}_{l]}\theta\right).
\end{split}
\end{equation}

Equation \eqref{derivative as Omega tensor} gives us the condition on $\theta $ for which a vacuum in the origin is found.
The matrices $\Gamma^{ijkl} $ are real, symmetric and traceless: consistently with their $\eta$-selfduality property, only 35 of them are independent and they form a basis of symmetric traceless matrices.
Hence \eqref{derivative as Omega tensor} reduces to
\begin{equation}
 2\theta^2\vert_{traceless}={\text{Tr}\theta}\,\theta\vert_{traceless}.
\end{equation}
We can combine this expression with \eqref{potential with only theta} and finally write the stationary point condition as
\begin{equation}\label{vacuum condition theta}
	2\,\theta^2 - \theta\,\text{Tr}\theta = V \, \mathbb 1.
\end{equation}
The computation of the minima and masses via (\ref{first derivative as isometry variation}) and (\ref{mass matrix as isometry variation}) leads to the same expressions.

We are now in position to determine the critical points at the origin of the scalar potentials generated by the $\theta$ tensors.
A first inspection of equation~\eqref{vacuum condition theta} shows that the condition to get an extremum of the scalar potential is invariant under similarity transformations
$\theta\rightarrow P\theta P^{-1}$.
This fact means that we can restrict our analysis to diagonal $\theta$ matrices, because (\ref{vacuum condition theta}) only poses conditions on the 8 eigenvalues of $\theta$.
Moreover, it is also independent on rescalings, so that we are left with 7 parameters:
\begin{equation}
 P\theta P^{-1}\propto\text{diag}(\lambda_1,\lambda_2,\lambda_3,\lambda_4,\lambda_5,\lambda_6,\lambda_7,1)\,.
\end{equation}
We can summarize this result as follows.
CSO($p,q,r $) gaugings are usually defined by a diagonal matrix $\theta^0_{p,q,r} $ with eigenvalues equal to $\pm 1$ or $0 $
\begin{equation}
 \theta^0_{p,q,r}=\text{diag}(\underbrace{1,\ldots,1}_p,
                \underbrace{-1,\ldots,-1}_q,
                \underbrace{0,\ldots,0}_r)
 \,,\quad
 p+q+r=8\,,
\end{equation}
from which the other embeddings are obtained by an SL(8) transformation $U $ that relates the generic symmetric matrix $\theta $  to $\theta^0_{p,q,r} $
\begin{equation*}
 \theta^0_{p,q,r}\quad \overset{SL(8)}\longrightarrow \quad \theta=U(\theta^0_{p,q,r})\,.
\end{equation*}
However, once more we can regard $U$ as acting on the coset representatives instead of the embedding tensor and take advantage of this fact to check for vacua that have non vanishing values of the scalar fields, with
\begin{equation*}
 L(\phi)=U^{-1}L(0)h^{-1}, \quad h\in SU(8).
\end{equation*}
For each diagonal $\theta$ that is a solution of \eqref{vacuum condition theta} there is always at least one transformation that maps $\theta^0_{p,q,r}$ into $\theta$, generated by the Cartan subalgebra of ${\mathfrak e}_{7(7)}$. 
This also fixes the normalization because these generators must be traceless. Indeed, these transformations are the diagonal $SL(8)$ matrices
\begin{equation}
 l_{A}{}^{B}=\text{diag}(l_1,l_2,\ldots,l_8)
 ,\qquad
 \prod_{i=1}^8l_i = 1.
\end{equation}
We can therefore identify $l_i = \sqrt{|\lambda_i|}$ when $\lambda_i\neq0$, though in the case $r\neq 0$ the form of $l_{A}{}^{B} $ is not fixed completely. 
In summary, since an $l$ transformation does not change the signature of $\theta$, nor the number of vanishing eigenvalues, we can identify the gauge group for each solution by counting the number of positive, negative and vanishing eigenvalues.

The residual gauge symmetry at the vacuum is given by the condition \eqref{residual sym condition at 0}
\begin{equation}
 g L(0) = L(0)h,
\end{equation}
where $g$ is generated by the gauge algebra defined by $\theta$ as opposed to $\theta^0_{p,q,r}$. 
Therefore we must require $g\in \text{SU}(8)$ which reduces to counting multiplicities of the nonvanishing eigenvalues of $\theta$ and associating a SO$(n)$ group each.

We can now provide the first results obtained with this method.
\begin{table}[ht]\renewcommand{\arraystretch}{1.2}\addtolength{\tabcolsep}{-1pt}%
\begin{center}
\rowcolors{1}{white}{gray!15}
\begin{tabular}{|c|l|c|}
\hline
	G$_{g}$		& $\vec \lambda$			& $\Lambda$\\
\hline\hline
	SO(8)		& $(1,1,1,1,1,1,1,1)$		&AdS\\
				& $(5,1,1,1,1,1,1,1)$		&AdS\\
	SO(3,5)		& $(-3,-3,-3,1,1,1,1,1) $	&dS\\
 	SO(4,4) 	& $(-1,-1,-1,-1,1,1,1,1)$ 	&dS\\
	CSO(2,0,6) 	& $(1,1,0,0,0,0,0,0)$ 		&Mink.\\
\hline
\end{tabular}
\end{center}
\caption{solutions of \eqref{vacuum condition theta} for $\theta=\text{diag}(\lambda_1,\ldots,\lambda_8)$, up to normalization. In the third column we indicate the type of vacuum arising from these solutions.}
\label{tab_eigenvalue solutions only theta}
\end{table}
The best initial approach is to simply solve \eqref{vacuum condition theta}  for a generic diagonal $\theta $ and then identify, if necessary, the transformation $l $ that leads from the standard form $\theta^0_{p,q,r} $ to the solution.
This is sufficient to reproduce 2 known vacua of the SO(8) theory and all known vacua of the CSO($p,q,r $) gaugings for non-zero $q$ and/or $r$.
For each solution we can now also provide the complete mass spectrum for the scalar fields and discuss the stability of the vacua (at least at quadratic order).
For most of these vacua this is a new result, unavailable with previous techniques.
In table \ref{tab_eigenvalue solutions only theta} we report the 5 combinations of eigenvalues that solve the vacuum condition.
These solutions are defined up to a rescaling, however this freedom is fixed by the requirement that $l\in  \,$SL(8).
In table \ref{tab_results for three known vacua at origin} we give the mass spectrum of the same vacua together with the values of the effective cosmological constant.

\begin{table}[hb]\renewcommand{\arraystretch}{1.3}\addtolength{\tabcolsep}{-1pt}%
\begin{center}
\rowcolors{1}{white}{gray!15}
\begin{tabular}{|c|c|l|}
\hline
	G$_{g}$		& $\Lambda$	& $m^2{}_{\text{ (multiplicity)}}$ \\
\hline\hline
	SO(8)		& $-6$ 	& $-\frac23{}_{(70)}$				\\
	SO(8) & $-2\times5^{3/4}$
	& $0_{(7)},\ 2{}_{(1)},\ -\frac45{}_{(27)},\ -\frac25{}_{(35)} $\\
	SO(3,5)		& $2\times3^{1/4}$
	& $0_{(15)},\ 
	4{}_{(5)},\ 
	-2{}_{(1)},\ 
	2{}_{(30)},$
	% \\  & &
	$\ \frac43{}_{(14)},\ 
	-\frac23{}_{(5)},$  \\
 	SO(4,4) 	& 2 	& $0_{(16)},\ 1_{(16)},\ 2_{(36)},\ -2_{(2)}$ \\
	CSO(2,0,6) 	& 0 	& $0_{(48)},\ \frac{1}{2}_{(20)},\ 2_{(2)}$ \\
\hline
\end{tabular}
\end{center}
\caption{Values of the effective cosmological constant and scalar masses in units of the cosmological constant (for $\Lambda \neq 0$).}
\label{tab_results for three known vacua at origin}
\end{table}

We can see that the SO(8) solution at the origin, corresponding to $\theta=\theta^0_{8,0,0}=\mathbb 1 $, is the well known maximally supersymmetric AdS vacuum of the SO(8) theory. Indeed, one sees immediately that $A_2 = 0$ and therefore 8 Killing spinors are preserved at the vacuum.
The Breitenlohner--Freedman bound $|m^2/\Lambda|\le3/4$ is of course satisfied and the vacuum is stable.

For the SO(4,4) gauging we have an unstable de Sitter vacuum for $\theta=-\mathbb 1_4\oplus\mathbb1_4$.
The 16 non-compact gauge symmetries are broken spontaneously at the vacuum.
Indeed, we can identify in the mass spectrum 16 vanishing eigenvalues corresponding to the Goldstone bosons.
The de Sitter vacuum of the SO(4,4) theory was known \cite{Hull:1984wa,Hull:1984ea} and part of its spectrum and its (in)stability was also discussed in \cite{Kallosh:2001gr,arXiv:0912.4440}.

For  $ \text{CSO}(2,0,6)$ we find a stable Minkowski vacuum at $\phi^{ijkl}=0 $.
There are 48 vanishing masses,
12 of which correspond to Goldstone bosons of the broken (non-compact) gauge symmetries.
Other flat directions can be identified with SL(8) transformations acting on $\theta=\theta^0_{2,0,6} $ of the form:
\begin{equation}
\left(\begin{array}{c|ccc}
		\alpha\mathbb1_2 & & & \\\hline
		 & & & \\
 		 & & M_{6\times6} & \\
		 & & & \\
\end{array}\right)
,\quad
\alpha=|{\text{det}M}|^{-1/2}\,,\quad
M\in \text{GL(6)} / \text{SO(6)}
\,.
\end{equation}
The requirement that $M$ does not belong to the compact subgroup SO(6) is motivated by the fact that such transformations are contained in the isotropy group SU(8) and therefore do not correspond to different points on the scalar manifold.
The GL(6) transformations account for additional 21 flat directions.
Of course, they also lead to a rescaling of $\theta $ by a factor $\alpha $, which is reflected in a multiplicative factor $\alpha^2 $ in the values of the masses indicated in table \ref{tab_results for three known vacua at origin}.
The remaining 15 null mass eigenvalues are associated with isometry transformations that are not contained in SL(8), and therefore correspond to shifts of the imaginary parts of the scalar fields.
This vacuum breaks supersymmetry completely, as we can easily see since $A_1\propto\mathbb1_8$, while we would expect it to have some vanishing eigenvalues if there were solutions to the Killing spinor equations.
Actually, it can be easily seen that such a model is also a special case of a more general flat group gauging, related to the Scherk--Schwarz reductions, where all the gravitino masses have been chosen equal \cite{companion}. 

The analysis of this vacuum suggests also an observation that goes beyond the application to this specific instance.
We know that Scherk--Schwarz models are models that have positive semidefinite potentials and hence admit only Minkowski minima.
This is translated into specific conditions on the embedding tensor that lead us to the same conclusion.
Among these, there is the fact that the variation of the potential with respect to the non-compact generators vanishes because of the non-trivial grading of the potential itself with respect to some of these generators.
It is interesting to see that this can be made into a general argument. 
Suppose that the whole embedding tensor has grading $c$ with respect to some non-compact generator $t \in \mathfrak e_{7(7)}$:
\begin{equation}
	(\delta_{t}\Theta)_M{}^\alpha = 
	 t_M{}^N \Theta_N{}^\alpha + \Theta_M{}^\beta t_\beta{}^\alpha
	 = c\, \Theta_M{}^\alpha.
\end{equation}
Then, all vacua admitted by the theory must be Minkowski.
Indeed the variation of the scalar potential with respect to such generator is proportional to the potential itself and therefore, for the first derivative to vanish, we need also the scalar potential to vanish:
\begin{equation}
\delta_{t} V \propto 
  (\delta_{t}\Theta)_M{}^\alpha \Theta_M{}^\beta (\delta_\alpha^\beta + 7 \eta_{\alpha\beta})
  = c\, V = 0 \quad \Leftrightarrow \quad V =   \Theta_M{}^\alpha \Theta_M{}^\beta (\delta_\alpha^\beta + 7 \eta_{\alpha\beta}) = 0.
\end{equation}
This is exactly what happens in a generic Scherk--Schwarz model, where the embedding tensor has a non-trivial grading with respect to an SO(1,1) generator, which is singled out in the electric E$_6$ basis \cite{Andrianopoli:2002mf}.

The other two solutions in table \ref{tab_eigenvalue solutions only theta} correspond to vacua that were found for non-vanishing values of some of the scalar fields.
We reproduce them here by performing an $l\in$ SL(8) transformation on the embedding tensor, as previously explained.
Since $\text{det}\,l=1 $, we need to normalize the eigenvalues indicated in the table, so that these two new vacua are associated with matrices
\begin{equation}
	\begin{array}{cc}
 \theta=5^{-1/8}(5\oplus\mathbb1_7),&\text{SO(8) vacuum,}\\[3mm]
 \theta=3^{-3/8}(-3\,\mathbb1_3\oplus\mathbb1_5), \qquad \phantom{.}&\text{SO(3,5) vacuum.}
	\end{array}
\end{equation}
The SO(8) vacuum \cite{Warner:1983vz,deWit:1983gs} is anti de Sitter and its residual symmetry group is SO(7).
For the 27 directions with negative mass eigenvalues $m^2=-8/5^{1/4}$, the Breitenlohner--Freedman bound is not respected, indeed one finds
$|m^2/\Lambda|=4/5>3/4$, which means that the vacuum is unstable.
Supersymmetry is broken completely.
The SO(3,5) vacuum is de Sitter and unstable \cite{Kallosh:2001gr,arXiv:0912.4440}.
The ratio of the negative mass eigenvalues with the effective cosmological constant is of order 1, as in the case of the SO(4,4) vacuum.
Therefore conditions for slow roll inflation are not satisfied by both these vacua.
The 15 vanishing masses corespond to Goldstone bosons.

We stress again that up to this point our analysis is based on the choice of the embedding tensor $\Theta_{M}{}^{\alpha}$ in the electric frame of SL$(8,\mathbb R) $, and that we are able to find vacua by using only SL(8) transformations.
There is a one-to-one correspondence between the generators of the  {non-compact}  isometry transformations and the scalar fields.
Indeed, with reference to the definitions \eqref{e7 gen in sl8 basis} and assuming a symmetric parametrization of the coset space, the 35 symmetric matrices $\Lambda_{A}{}^{B} $ that generate SL(8)/SO(8) correspond to the real parts of the scalar fields, while the selfdual tensors $\Sigma_{ABCD} $ correspond to the imaginary parts.
We can therefore state that these are all the vacua where the scalar fields take real values in the standard parameterization of the CSO($p,q,r$) gaugings.
Other vacua are known for the SO(8) theory, however the scalar fields generally take complex values at the critical points.
In order to reproduce these vacua, we would have to consider all non-compact isometries of the coset space E$_{7(7)}$/SU(8).

% subsection _theta_neq_0_ (end)

\subsection{$\theta \neq 0$ and $\xi \neq 0$} % (fold)
\label{sub:_theta_neq_0_and_xi_neq_0_}

There is another similar setup that not only allows us to reproduce some other vacua of the SO(8) theory, but also to reveal new stationary points and new gauge groups embedded in SL(8).
We begin our discussion with the study of the SO(8) theory. We would like to reproduce vacua where the scalar fields take purely imaginary values, which translates into considering isometry transformations generated by the 35 sefldual tensors $\Sigma_{ABCD}$. 

With reference to \eqref{e7 gen in sl8 basis}, one can identify a $\mathfrak{su}$(8) and two $\mathfrak{sl}$(8) subalgebras of ${\mathfrak e}_{7(7)}$ \cite{LPTENS 79/6}. 
The first $\mathfrak{sl}$(8) is given in terms of traceless matrices $\Lambda$, while $\mathfrak{su}$(8) is constructed taking $\Lambda=-\Lambda^T$ and adding the 35 anti-selfdual tensors $\Sigma_{ABCD}$. 
A second $\mathfrak{sl}$(8) algebra can be defined in terms of $\Lambda=-\Lambda^T$ and selfdual $\Sigma_{ABCD}$. 
We will denote the group generated by the latter algebra SL$'(8)$, to distinguish it from the previous one. We stress that all three groups share the same SO(8) generated by antisymmetric $\Lambda$'s. We see that the non-compact part of SL$'$(8) is associated to the pseudoscalar fields. 
We would like therefore to rewrite the generators and the embedding tensor in an SL$'(8)$ covariant form.
This is achieved in a manner similar to what we did in \eqref{cayley matrix and complex basis}: we introduce a new set of chiral $\Gamma$ matrices, $\Gamma'{}_{ab}^{AB}$ that interpolate between SL(8) and SL$'$(8) indices (upper- and lower-case respectively). 
We must choose the selfduality relations 
\begin{equation}\label{Gamma' selfduality conds}
	[\Gamma'_{ABCD}]_a{}^b = [\star\Gamma'_{ABCD}]_a{}^b,\qquad [\Gamma'_{abcd}]_A{}^B = [\star\Gamma'_{abcd}]_A{}^B,
\end{equation}
in order to obtain the correct change of basis. We stress that $\Gamma'$ matrices can be taken to be real, which allows us to define the symplectic rotation
\begin{equation}\label{sl -> sl' rotation}
	R=\frac{1}{4\sqrt2}\begin{pmatrix}
		\Gamma_{ab}^{AB} & \Gamma_{abAB} \\
		-\Gamma^{abAB} & \Gamma^{ab}_{AB} \\
\end{pmatrix}.
\end{equation}
Generators $R\, t_\alpha\, R^T$ are formally identical to \eqref{e7 gen in sl8 basis}, with upper-case indices replaced by lower-case ones. The old SL(8) group is now generated by $\Lambda_a{}^b= -\Lambda_b{}^a$ and selfdual $\Sigma_{abcd}$.
The gauge algebra $\mathfrak{so}(8)$ is not affected by $R$.

We can now apply the same trasformation to the embedding tensor, assuming $\theta_{AB} = \delta_{AB}$. The calculation is very similar to the one of the fermion shifts. It is evident from \eqref{sl -> sl' rotation} that $\Theta_M{}^\alpha$ in the new basis will \emph{not} be electric. Namely, we expect $\Theta^\Lambda \neq 0$. Indeed, we find
\begin{equation}\label{gauge struct constants theta' and xi'}
	X_{ab}=\begin{pmatrix}
		-2\delta_{[a}^{[e} \theta'{}_{b][c}^{\vphantom{[]}}\delta_{d]}^{f]} &  \\
		 & 2\delta_{[a}^{[c} \theta'{}_{b][e}^{\vphantom{[]}}\delta_{f]}^{d]} \\
\end{pmatrix},
\quad
X^{ab}=\begin{pmatrix}
		-2\delta_{[c}^{[a} \xi'{}^{b][e}_{\vphantom{[]}}\delta_{d]}^{f]} &  \\
		 & 2\delta_{[e}^{[a} \xi'{}^{b][c}_{\vphantom{[]}}\delta_{f]}^{d]} \\
\end{pmatrix},
\end{equation}
where for the moment $\theta'=\xi'=\mathbb1_8$.
If we did not take $\theta_{AB}=\delta_{AB}$, extra terms proportional to $[\Gamma'_{a}{}^{bcd}]^{AB}\theta_{AB}$ would have appeared, corresponding to the $\bf 420$ and $\bf 420'$ $\mathfrak{sl}'(8)$ representations.
We have already discussed the quadratic constraint for fluxes in the $\bf36$ and $\bf36'$, which reduces to $\xi'\propto\theta'^{-1}$.
This is of course automatically satisfied and we are now free to apply SL$'(8)$ transformations to this setup, which translates into taking an arbitrary symmetric, positive-definite and unimodular matrix $\theta'$, together with its inverse $\xi'=\theta'^{-1}$. 
A direct calculation of the potential gives
\begin{equation}\label{V with theta' and xi'}
	V(0,\theta',\xi') = \frac{1}{8}\text{Tr} \,(\theta'^2)-\frac{1}{16}(\text{Tr}\,\theta')^2
                +\frac{1}{8}\text{Tr}\,(\xi'^2)-\frac{1}{16}(\text{Tr}\,\xi')^2.
\end{equation}

Taking derivatives with respect to the pseudoscalar fields is equivalent to taking infinitesimal non-compact SL$'(8)$ variations of $V(0,\theta',\xi')$. One obtains the stationary point condition
\begin{equation}
 2\text{Tr}\left(\Lambda(\theta'^2 - \xi'^{2})\right) - \text{Tr}\left(\Lambda(\theta'\text{Tr}\theta'-\xi'\text{Tr}\xi') \right) = 0, \quad \Lambda\in \mathfrak{sl}'(8).
\end{equation}
We remind that $\xi'$ transforms in the conjugate representation with respect to $\theta'$.
Now, since $\Lambda$ can be any symmetric traceless matrix, we must just require the rest of the expression to vanish up to some term proportional to the identity:
\begin{equation}\label{vacuum condition theta' and xi'}
	2(\theta'^2 - \xi'^{2}) - (\theta'\text{Tr}\theta'-\xi'\text{Tr}\xi') \propto \mathbb 1_8,
\end{equation}
In order to prove that this is the correct vacuum condition, we need to check that SL(8) variations vanish identically. 
This can be done explicitly from \eqref{first derivative as isometry variation}, noting that no mixed terms with $\theta'$ and $\xi'$ appear, so that one can separate the structure constants in two pieces and easily recover the final result\footnote{One should also check that no contributions come from the $\bf420$ and $\bf420'$ terms. This is guaranteed by direct calculation of $V(\phi=0)$ in the general case, which shows that $ V \sim (\bf36)^2 + (\bf36')^2 + (\bf420)^2 + (\bf420')^2 $ and we set the last two contributions to zero from the beginning.}.

The vacuum condition \eqref{vacuum condition theta' and xi'} is again invariant under similarity transformations. 
We can therefore restrict our attention to diagonal matrices. 
By fixing $\det \theta' = 1$, we end up with only 7 parameters. 
We stress that a priori we are considering a setup where 72 parameters of the embedding tensor are turned on, although only half of them are independent after we impose the quadratic constraint. Up to an irrelevant normalization, this mirrors the fact that we are reproducing vacua where 35 (pseudo) scalar fields are allowed to take non-vanishing values. The fact that we can restrict to diagonal $\theta'$ and $\xi'$ without loss of generality is therefore an important simplification.
Unfortunately, even after such simplifications, the equation determining the vacua is not anymore quadratic in the $\theta'$ parameters, because of the presence of the inverse matrix $\xi' \propto \theta^{\prime-1}$. 
However, one can still find analytic solutions by varying 1 or 2 eigenvalues. 
We chose
\begin{equation}
	\theta'_{ab} = (r\, s)^{-1/8}\text{diag}(1,1,1,1,1,1,r,s)
\end{equation}
and in addition to the $N=8$ AdS vacuum at $r=s=1$, we found two more solutions
\begin{itemize}
%	\item $r = s = 1$, which is again the $N=8$ AdS vacuum with $\Lambda=-6$.
	\item[i.] $r=1,\, s=(7-3\sqrt5)/2$, which is the SO$(7)_-$ vacuum\footnote{The $\pm$ symbol differentiates the two SO(7) vacua found so far in the $N=8$ theory. These are related by a ${\mathbb Z}_2$ symmetry.} described in \cite{Warner:1983vz,deWit:1983gs}, with $\Lambda/g^2=-\frac{25\sqrt5}{8}$;
	\item[ii.] $r=3+2\sqrt2,\,s=1/r$, which is the $SU(4)_-\simeq SO(6)$ vacuum found in \cite{Warner:1983vz}, with $\Lambda/g^2=-8$.
\end{itemize}
We will discuss further these and other results below.

%%%%%%%%%%%%%%%%%%%%% %%%%%%%%%%%%%%% %%%%%%%%%%%%%%% %%%%%%%%%%

Inspired by this construction, we can now relax the conditions on $\theta'$ and $\xi'$, in order to identify other vacua and other gauge theories.
From now on, we drop the $'$ from the notation, since the SL(8) and SL$'(8)$ bases are formally identical.
Moreover, we will not require $\theta$ and $\xi$ to be positive definite nor impose any particular value to their determinant, since normalizations can be worked out later.
All relevant formulas in this section still apply, in particular \eqref{gauge struct constants theta' and xi'}, \eqref{V with theta' and xi'} and \eqref{vacuum condition theta' and xi'}.
We therefore consider the general solution of the quadratic constraint for these fluxes, namely
\begin{equation}
 \xi \propto \theta^{-1}\quad\text{or}\quad \theta\,\xi=0.
\end{equation}
In the first case, the resulting gauge algebra is still $\mathfrak{so}$($p,q$), with $p+q=8$, however the gauge connection now also involves the ``magnetic'' vector fields $A_{\mu\,AB}$.
Contractions to $\mathfrak{cso}$($p,q,r$) are no more allowed, because they would violate the quadratic constraint. 
We can, however, consider the case $\theta\,\xi = 0$.
Since both $\theta$ and $\xi$ must have a non-empty kernel, the gauge algebras turn out to be ``superpositions'' of two $\mathfrak{cso}(p,q,r$) ones, where the two semisimple factors commute with each other (because of the quadratic constraint), while some of the nilpotent generators are in common between the groups and others add up to form a bigger abelian algebra. This structure can be guessed by inspecting \eqref{sl8embedding}, \eqref{sl8 embedding of xi} and \eqref{gauge struct constants theta' and xi'}. 
The generic form of these gaugings is
\begin{equation}
 \left(\text{SO}(p,q) \times \text{SO}(p',q')\right) \ltimes T^{ (8-r)r + (8-r')(r' + r -8) }
\end{equation}
where $p+q+r=p'+q'+r'=8$. They can be defined as different contractions of SO($p+p',8-p-p'$).

The vacuum condition for this setup is still given by \eqref{vacuum condition theta' and xi'}. 
Again, finding all solutions analytically is difficult, but several new results can already be obtained by varying only a subset of the eigenvalues of $\theta$ and $\xi$.
We summarize our results in Table \ref{tab_riassunto}.

\begin{table}[htb]\renewcommand{\arraystretch}{1.4}\addtolength{\tabcolsep}{0pt}
\begin{center}
\rowcolors{3}{white}{gray!15}
\begin{tabular}{|c|c|l|c|}
\hline
\multirow{2}{*}{$\#$} &\multirow{2}{*}{Gauging} &\multirow{2}{*}{$\Theta_M{}^\alpha$}	&\multirow{2}{*}{$\Lambda$}\\
&&&\\\hline\hline
i	 &SO(8) 									& $ \theta = \xi^{-1} = a \oplus \mathbb 1_7$ 																		& AdS \\
ii	 &SO(8) 									& $ \theta = \xi^{-1} = (a,1/a)\oplus\mathbb 1_6$																	& AdS \\
iii	 &SO(7,1) 									& $ \theta = \xi = -1 \oplus \mathbb 1_7$ 																			& AdS \\
iv	 &SO(7,1) 									& $ \theta = \xi^{-1} = (b_1, b_2)\oplus\mathbb 1_6$ 																& AdS \\
v	 &SO(7,1) 									& $ \theta = \xi^{-1} = -1\oplus\mathbb 1_5\oplus(c,1/c)$ 															& AdS \\
vi	 &SO(6,2) $\simeq$ SO$^*(8)$					& $ \theta = \xi = (-\mathbb 1_2 \oplus \mathbb 1_6)/2\sqrt2$ 														& Mink \\
vii	 &SO(5,3) 									& $ \theta = \xi = -\mathbb 1_3 \oplus \mathbb 1_5$																	& dS  \\
viii &SO(7) $\ltimes\ T^7$						& $ \theta =	0 \oplus \mathbb 1_7,$\  $\xi=\sqrt5\oplus\mathbb 0_7	$											& AdS \\
ix	 &SO(7) $\ltimes\ T^7$						& $ \theta = 4 \oplus \mathbb 1_6 \oplus 0,$\  	$\xi=\mathbb 0_7 \oplus 2\sqrt2$ 									& AdS \\
x	 &SO(6) $\times$ SO(1,1) $\ltimes\ T^{12}$	& $ \theta = \mathbb 1_6 \oplus \mathbb 0_2,\ $ $\xi=\mathbb 0_6 \oplus (\sqrt2,-\sqrt2)$	 						& AdS \\
xi	 &SO(6) $\times$ SO(1,1) $\ltimes\ T^{12}$	& $ \theta = 3 \oplus \mathbb 1_5 \oplus \mathbb 0_2,\ $ 	$\xi=\mathbb 0_6 \oplus (\sqrt3, -\sqrt3)$ 				& AdS \\
xii	 &SO(4) $\times$ SO(2,2) $\ltimes\ T^{16}$	& $ \theta = \mathbb 1_4 \oplus \mathbb 0_4,\ $	 $\xi=\mathbb 0_4 \oplus\mathbb 1_2\oplus -\mathbb 1_2$ 			& Mink \\
xiii &SO(2)$^2$ $\ltimes\ T^{20}$				& $ \theta = (\mathbb 1_2\oplus\mathbb 0_6)\sqrt2,\ $ $\xi=(\mathbb 0_2 \oplus\mathbb 1_2\oplus\mathbb 0_4)\sqrt2$ 	& Mink \\\hline
\end{tabular}
\caption{%
Vacuum solutions found for $\theta ,\xi\neq0$. The SO(8) case is also included here, although with a different normalization.
Minkowski solutions are normalized to easily compare their mass spectra (see Table \ref{tab_mass_spectra}).
$a=3+2\sqrt2$; $b_1=1/2(-1+~\sqrt2)(-1+\sqrt5)$, $b_2=-1/2(1+~\sqrt2)(-1+\sqrt5)$;
$c=2+\sqrt3$.}
\label{tab_riassunto}
\end{center}
\end{table}

\begin{table}[tb]\renewcommand{\arraystretch}{1.3}\addtolength{\tabcolsep}{-1pt}
\begin{center}
\label{tab_mass_spectra}
\begin{tabular}{|c|l|l|c|c|}
 \hline
$\#$&$G_{gauge}$ 						& $G_{res}$				& $\Lambda$	& $m^2{}^{\ (multipl.)}$									\\
\hline
\hline
\rowcolor{gray!15}vi&SO$(2,6)$						& SO(2) $\times$ SO(6)	& 	& 	\\
\rowcolor{gray!15}&CSO$(2,0,6)$ 						& SO$(2)$ 				&	\multirow{-2}{*}{Mink}			& 		\multirow{-2}{*}{$2^{(2)}, \frac12^{(20)}, 0^{(48)}$}					\\
xii&SO$(4)\times$ SO$(2,2)\ltimes T^{16}$ & SO(2)$^2\times$ SO(4) & \multirow{2}{*}{Mink}		&  \multirow{2}{*}{$4^{(4)}, 2^{(12)}, 1^{(16)}, 0^{(38)}$}					\\
xiii&SO$(2)^2\ltimes T^{20}$ 			& SO$(2)^2$				&			&										\\
\rowcolor{gray!15}i&SO$(8)$								& 				&		& 						\\
\rowcolor{gray!15}iii&SO$(7,1)$							&		SO$(7)$				&	 AdS		&			$2^{(1)}, -\frac45^{(27)}, -\frac25^{(35)}, 0^{(7)}$											\\
\rowcolor{gray!15}viii&SO$(7)\ltimes T^7$					&						&			&										\\
ii&SO$(8)$								& \multirow{4}{*}{SO$(6)$}			&  \multirow{4}{*}{AdS}		& \multirow{4}{*}{ $2^{(2)}, -1^{(20)}, -\frac14^{(20)}, 0^{(28)}$}				\\
iv&SO$(7,1)$							&						&			&															\\
ix&SO$(7)\ltimes T^7$					&						&			&															\\
x&SO$(6)\times$ SO$(1,1)\ltimes T^{12}$	&						&			&										\\
\rowcolor{gray!15}v&SO$(7,1)$							&  			&  		&  				\\
\rowcolor{gray!15}xi&SO$(6)\times$ SO$(1,1)\ltimes T^{12}$	&	\multirow{-2}{*}{SO$(5)$}						&	\multirow{-2}{*}{AdS}		&											\multirow{-2}{*}{$ 2^{(3)}, -\frac43^{(14)}, \frac23^{(5)}, 0^{(48)}$}\\
vii&SO$(3,5)$							& SO(3) $\times$ SO(5)	& dS		& $ -2^{(1)}, 4^{(5)}, 2^{(30)}, \frac43^{(14)}, -\frac23^{(5)}, 0^{(15)}$	\\[2mm]
\hline
\end{tabular}
\caption{%
mass spectra and residual symmetries for the new vacua. Known solutions of CSO$(2,0,6)$ and SO$(8)$ theories are given for reference. When $\Lambda\neq0$, masses are normalized with respect to it.%
}
\end{center}
\end{table}

Inspecting Table \ref{tab_riassunto}, we see that $\theta$ and $\xi$ have no fixed normalization.
This contrasts with the analysis presented previously and therefore it deserves an explanation.
When only $\theta \neq 0$, we can obviously change the values of $\theta$ to move on the parameter space associated to the moduli space, but we can also rescale the gauge coupling constant to define equivalence classes of gaugings.
In particular, we have seen that once we normalize the value of the cosmological constant for a specific vacuum of a given gauging, we can keep consistent normalizations by using rescalings that do not change the determinant of $\theta$.
On the other hand, when $\xi=c\,\theta^{-1}$ for $c\neq0$, there are three parameters that can be tuned: the proportionality constant $c$, the determinant of $\theta$, and the gauge coupling constant $g$. 
Moreover, a rescaling of the coupling constant acts in the same way on $\theta$ and $\xi$, while a rescaling of $\det\theta$ acts inversely on $\xi$. 
Introducing equivalence classes of gaugings for different parameterization is therefore more subtle.
We have seen previously that $\det\theta$ is preserved by isometry transformations. 
However, the quadratic constraint ties $c$ and $\det\theta$ with each other, so that, by rescaling $g$, one could either fix $\det\theta=\pm1$ and keep $c\neq0$ or alternatively fix $c=1$ and let $\det\theta$ to take any non-vanishing value. 
Actually, once the normalization of the value of the cosmological constant of a given vacuum for a specific gauging has been fixed, other vacua will keep the same normalization with respect to that only for fixed combinations of the various rescalings, which, however, generically depend on the gauging.
For instance, in addition to the vacua in Table \ref{tab_riassunto}, we also find an SO(4,4) vacuum in this class, whose cosmological constant can be normalized to the same value as the one of the vacuum obtained with only $\theta \neq 0$.
In fact we can show that this new vacuum constructed with both $\theta \neq 0$ and $\xi \neq 0$ is equivalent to the previous one.
This can be done by performing a symplectic rotation of a form analogous to the one used for the SO(8) gauging, namely (\ref{sl -> sl' rotation}). 
The relevant matrix can be constructed using a real representation for the $\Gamma_{AB}^{ab}$ matrices, with $(\Gamma_A)^{ab}$ and $(\Gamma^a)_{AB}$ elements in Cliff(4,4), satisfying self-duality conditions (\ref{Gamma' selfduality conds}) where the indices are raised and lowered using $\theta$.
Unfortunately, it is not easy to envisage the explicit form of a similar transformation for the other cases, where the corresponding $\Gamma$ matrices cannot be chosen to be real and the transformation itself must include a non-compact element of E$_{7(7)}$.
This is the reason why we simply decided to fix $c=1$ in all the remaining examples and report the vacua with arbitrary $\det \theta$.
We made an exception for the Minkowski solutions (vi) and (xiii), where a different normalization is chosen in order to better compare the mass spectra with other vacua.

We have checked explictly the residual supersymmetry, gauge symmetry and stability of these vacuum solutions. 
All vacua break supersymmetry completely, and the AdS ones do not respect the Breitenlohner--Freedman bound. 
However, Minkowski solutions are stable at the quadratic level.  
Table \ref{tab_mass_spectra} summarizes these results.

We were also able to identify some flat directions, not corresponding to Goldstone bosons, for the SO(2,6) vacuum (vi) and for the SO(4)~$\times$~SO(2,2)~$\ltimes T^{16}$ (xii) Minkowski solutions.
In the first case, the following parametrization keeps $V=0$:
\begin{equation}\label{flat directions of SO(2,6)}
 \theta=\xi^{-1}=\text{diag}\left(-{r  s },-{r  s },\frac{s }{t  r },\frac{s }{t  r }, \frac u s , \frac u s ,\frac{t }{u s},\frac{t }{u s} \right),
\end{equation}
while for SO(4) $\times$ SO(2,2) $\ltimes T^{16}$
\begin{equation}
 \theta=\left(u ,u ,\frac{r  t }{u },\frac{r  t }{u },0,0,0,0\right),\quad
 \xi=\left(0,0,0,0,t  s ,t  s ,-\frac{r }{s },-\frac{r }{s }\right).
\end{equation}
Also for SO(2) $\times$ SO(2) $\ltimes T^{20}$, $\theta$ and $\xi$ can be rescaled independently without affecting the value of the potential.
An interesting fact about these flat directions is that they allow to interpolate between the SO(2,6) theory and the other two groups, when taking a singular limit of the isometry parameters. These non-semisimple groups, together with CSO(2,0,6), are indeed different contractions of SO(2,6). 
For example, with reference to \eqref{flat directions of SO(2,6)}:
\begin{itemize}
 \item $g\sim s\rightarrow0$ reproduces the contraction SO(2,6) $\rightarrow$ SO(4) $\times$ SO(2,2) $\ltimes T^{16}$,
 \item $g\sim u\rightarrow0$ reproduces SO(2,6) $\rightarrow$ SO(2) $\times$ SO(2) $\ltimes T^{20}$,
 \item $g\sim u s , s\sim u\rightarrow0$ reproduces SO(2,6) $\rightarrow$ CSO(2,0,6),
\end{itemize}
where $g$ is the gauge coupling constant, which must also take a singular limit in order to reproduce the correct Lie algebra contraction.

% subsection _theta_neq_0_and_xi_neq_0_ (end)

\subsection{On the spectra of the SO(6,2) and CSO(2,0,6) models} % (fold)
\label{sub:On the spectra of the SO(6,2) and CSO(2,0,6) models}

A first look at Table~\ref{tab_mass_spectra} reveals that the critical points of the SO(6,2) and CSO(2,0,6) gaugings with maximal residual symmetry group have the same mass spectra for the scalar fields.
This is also true for other gauge groups and vacua that share the same residual symmetry group.

% \begin{wrapfigure}{r}{0.55\textwidth} 
\begin{figure}[htb]
% \vspace{-9mm}
\begin{center}
	\includegraphics[scale=.27]{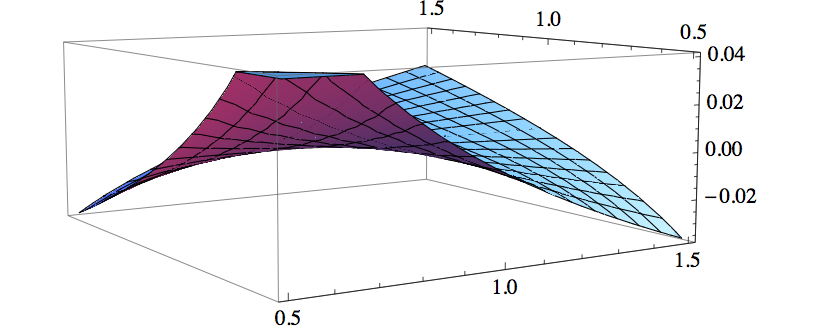}
	\includegraphics[scale=.27]{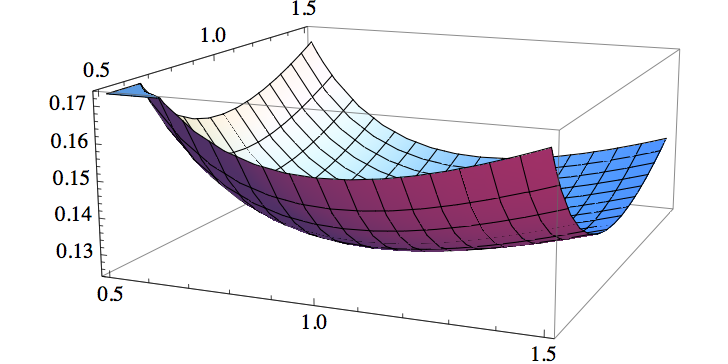} 
	\includegraphics[scale=.18]{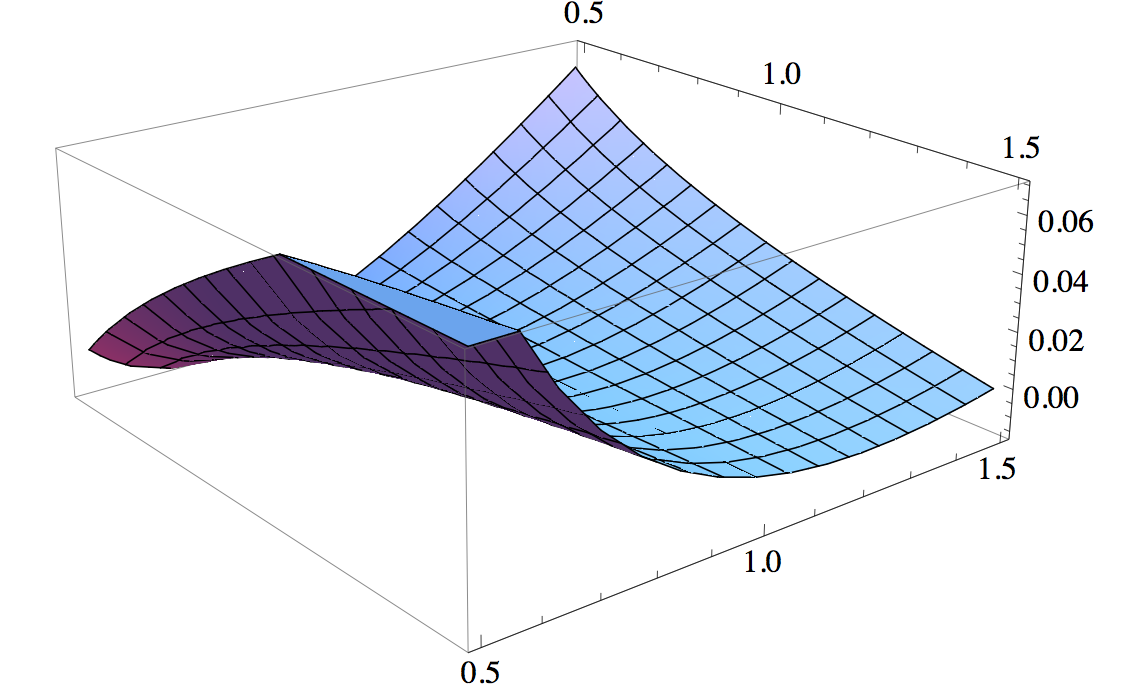}
\end{center}
% \vspace{-7mm}
\caption{\small \emph{Values of the masses of 3 of the scalar fields of the SO(6,2) spectrum as a function of two of the moduli fields. The first and the last plot are referred to massless scalars at the maximally symmetric critical point. It can be seen that in some directions the masses become negative and the potential unstable.}} \label{massplot}
\end{figure}
% \end{wrapfigure}

Of course, one could ask if this is a feature only of the scalar field masses or also of the other fields.
By applying our results for the fermion shifts to the mass formulae in \cite{LeDiffon:2011wt}, we computed also the masses of the other fields of the SO(6,2) and CSO(2,0,6) models and we actually found that the spectra coincide and agree with the spectrum of the Scherk--Schwarz models with 4 identical parameters, which is represented by the CSO(2,0,6) gauging.
In detail, besides the massless graviton, we have 8 gravitini with the same mass $m = \frac{1}{2\sqrt2}$, we have 8 massless spin 1/2 fields (corresponding to the 8 goldstini), 8 spin 1/2 fields with mass $3m$ and 40 spin 1/2 fields with mass $m$. 
We also have 16 massless vector fields and 12 vector fields with mass squared $4m^2$.

At this point one may think that the two models agree beyond the quadratic order, but they actually do not.
A more careful inspection of the spectrum reveals a very interesting dependence on the moduli fields.
While in the CSO(2,0,6) case, the dependence on the value of the moduli fields is such that one simply has an overall rescaling of the full spectrum, for the SO(6,2) model, the dependence is more complicated.
While moving along some of the directions dictated by such moduli fields one obtains a further breaking of the residual gauge symmetry and also the spectrum changes accordingly.
Actually, we can also see that going beyond the quadratic level there are some massless fields of the spectrum that become tachionic, rendering the vacuum unstable.
We provided in Figure \ref{massplot} an example of this phenomenon.
One can see from the plots that while moving in some directions in the moduli space one can further lift some of the flat directions existing at the origin of the moduli space, moving in other directions may lower the value of the masses of such scalar fields below zero.

Combining this information with the previous observation that at the boundary of the moduli space one recovers one model from the other, we can see that a very interesting structure emerges that we hope we will be able to analyze in more detail in the future.

% subsection On the spectra of the SO(6,2) and CSO(2,0,6) models (end)

% section vacua_of_the_gauged_theory (end)

\section{Comments} % (fold)
\label{sec:comments}

We have shown that by combining the embedding tensor formalism and the fact that the scalar manifold of $N=8$ supergravity is a coset space, we can reduce the conditions to obtain vacua of the gauged theory to linear and quadratic constraints on the embedding tensor.
Using this technique we easily reproduced many of the known vacua and produced a dozen new ones, while computing the full mass spectra for the scalar fields. All these results are analytical.

Actually, an interesting fact emerges by the analysis of such spectra. 
All the vacua that have the same residual gauge symmetry also have the same mass spectrum (normalized with respect to the cosmological constant).
While it can be expected that the scalar fields arrange themselves in multiplets of the residual symmetry and that therefore different vacua may show similar degeneracies, the fact that their exact mass value also coincides is somehow unexpected and deserves an explanation.

We should also note that the Minkowski vacua we found have moduli fields, whose expectation values tunes the values of the masses of the other scalar fields.
These are the $N=8$ analog of the $N=1$ no-scale models, where the supersymmetry breaking scale and the masses of the scalar fields depend on a sliding parameter determined by the vacuum expectation value of some modulus.
It is also a common feature of Scherk--Schwarz reductions, where all the masses depend on specific combinations of up to 4 parameters, but whose overall scale is fixed by some of the remaining moduli fields.
Our analysis shows that this feature is related to the fact that all the corresponding embedding tensors are charged with respect to some non-compact symmetry, whose corresponding isometry direction determines the scalar associated to such sliding scale.
In particular, both the CSO(2,0,6) and the SO(2) $\times$ SO(2) $\ltimes\ T^{20}$ models have an embedding tensor with a definite non-zero grading with respect to an SO(1,1) generator and hence fall directly into this class.
On the other hand also the SO(2,6) and the SO(4) $\times$ SO(2,2) $\ltimes\ T^{16}$ models have moduli dependent masses, although their embedding tensor is not of this type (and in fact the scalar potential is not positive definite and the dependence of the mass spectrum on the moduli fields is more complicated).

Another interesting fact was pointed out at the end of the previous section.
Once one has fixed a gauge group and found a vacuum, by moving on the residual moduli space towards the boundary one may reach a point that corresponds to a different choice of the embedding tensor and hence of the gauge group (up to a redefinition of the gauge coupling constant in the singular limit).
If one considers these models as products of (non-geometric) flux compactifications of some higher-dimensional theory, this fact suggests the existence of a non-trivial structure of the moduli space, with possible transitions between different geometries and compactification schemes.

Also in the vein of regarding at our 4-dimensional models as products of some compactification scheme, it is interesting to point out that the vacuum (x) has a gauge group whose semi simple part becomes electric in the standard frame following from reductions from 5 dimensions.
In such a frame the SO(6) gauge group could be naturally interpreted as the gauging resulting from the reduction on $S^5$, while the reduction from 5 to 4 dimensions must be performed in a way that produces an additional gauging of an SO(1,1) dilatation symmetry.
This suggests the existence of an $AdS_4 \times {\mathbb H}^1 \times S^5$ vacuum of type IIB supergravity.

We decided not to investigate further the higher-dimensional origin of our models, but we refer the reader to a companion paper where the relation between $N=8$ gauged supergravity and U-dual reductions of M-theory is discussed in detail \cite{companion}.

There are many obvious lines of development of our project.
The first one is to apply more systematically our approach in order to complete the classification of the vacua of $N=8$ $SO(8)$ gauged supergravity.
This gauged supergravity is of primary importance also for its relation to the models of M2-branes.
The technique presented here could be applied to such a theory after we have identified the duality orbit of SO(8) in E$_{7(7)}$.
Once this is done, one has fixed the general form of the embedding tensor related to such models and one can therefore apply our technique to this restricted $\Theta$.

Another interesting line of development is given by a better analysis of the criteria needed to obtain vacua with a positive cosmological constant.
For instance, we have seen that our analysis implies that having non-compact generators involved in the gauging is a necessary requirements for semisimple gaugings involving all 28 vector fields that couple to canonically normalized generators.
More in general, compact gaugings lower the value of the cosmological constant by a factor proportional to the group metric.
Although we could not extend this argument more in general, because embeddings with non canonical normalizations may appear, it is possible that by using also the information coming from the critical point conditions more strict relations between the parameters may appear, providing criteria with a wider applicability.
Once such constraints have been better understood one could also produce more example and have a final word on the stability of such vacua in extended supergravity.

Finally, the long-term goal of exhausting and classifying all possible gaugings and their vacua could be obtained by implementing the algorithm presented here on a computer, possibly combining this kind of analysis with advanced numerical techniques.

% section comments (end)

%%%%%%%%%%%%%%%%%%%%%%%%%%%%%%%%%%%%%%%%%%%%%%%%%%%%%%%%%%%%%%

\bigskip
\section*{Acknowledgments}

\noindent We would like to thank L.~Andrianopoli, M.~Bianchi, G.~Bossard, A.~Marrani, M.~Trigiante and especially F.~Catino, G.~Villadoro and F.~Zwirner for helpful discussions.
This work is supported in part by the ERC Advanced Grant no. 226455, \textit{``Supersymmetry, Quantum Gravity and Gauge Fields''} (\textit{SUPERFIELDS}), by the European Programme UNILHC (contract PITN-GA-2009-237920), by the Padova University Project CPDA105015/10 and by the MIUR-PRIN contract 2009-KHZKRX.
% grant RBFR10QS5J \emph{``String Theory and Fundamental Interactions''}.

%%%%%%%%%%%%%%%%%%%%%%%%%%%%%%%%%%%%%%%%%%%%%%%%%%%%%%%%%%%%%%

%%%%%%%%%%%%%%%%%%%%%%%%%%%%%%%%%%%%%%%%%%%%%%%%%%%%%%%%%%%%%%

\end{document}